\documentclass[%
 preprint,
 amsmath,amssymb,
 aps,
]{revtex4-1}

\usepackage[T1]{fontenc} 
\usepackage[english]{babel} 
\usepackage{amsmath,amsfonts,amsthm} 
\usepackage{graphicx}   
\usepackage{verbatim}   
\usepackage{color}      
\usepackage{subfigure}  
\usepackage{hyperref}   
\usepackage{graphicx}
\usepackage{dcolumn}
\usepackage{bm}
\usepackage{ulem}
\usepackage{xcolor}
\numberwithin{figure}{section} 
\numberwithin{table}{section} 

\newcommand*{\abb}[0]{$\text{(MC)}^2$~}

\begin{document}

\preprint{APS/123-QED}

\title{ Efficient determination of solid-state phase equilibrium with the Mutli-Cell Monte Carlo method}

\author{Edwin Antillon }\email{edwin.antillon@gmail.com}
\author{Maryam Ghazisaeidi}%
 \email{ghazisaeidi.1@osu.edu}
\affiliation{%
  Department of Materials Science at The Ohio State University
}%


\date{\today}

\begin{abstract}

Building on our previously introduced Multi-cell Monte Carlo \abb method for modeling phase coexistence,  this paper provides important improvements for efficient determination of phase equilibria in solids. 
The \abb method uses multiple cells, representing possible phases. Mass transfer between cells is modeled virtually by solving the mass balance equation after the composition of each cell is changed arbitrarily. However, searching for the minimum free energy during this process poses a practical problem. The solution to the mass balance equation is not unique away from equilibrium and consequently the algorithm is in risk of getting trapped in nonequilibrium solutions. Therefore, a proper stopping condition for \abb is currently lacking. In this work, we introduce a consistency check via a predictor-corrector algorithm to penalize solutions that do not satisfy a necessary condition for equivalence of chemical potentials and steer the system towards finding equilibrium. The most general acceptance criteria for \abb is derived starting from the isothermic-isobaric Gibbs Ensemble for mixtures. Using this ensemble, translational MC moves are added to include 
vibrational excitations as well as volume MC moves to ensure the condition of constant pressure and temperature entirely with a MC approach, without relying on any other method for relaxation of these degrees of freedom.  As a proof of concept the method is applied to two binary alloys with  miscibility gaps and a model quaternary alloy, using classical interatomic potentials.

\end{abstract}
\maketitle 

\date{\normalsize\today} 

\maketitle 

\newpage
\section{Motivation}
Knowing the stability of phases is of great importance for materials research and development. A precise determination of phase coexistence using atomistic simulations poses great challenges due to the long time scales needed to reach thermodynamic equilibrium. 
Unlike molecular dynamics methods which generate successive atomic configurations by integrating Newton's equation of motion, Monte Carlo (MC) methods are able to sample various arrangements of atoms that depend only on the previous state~\cite{dubbeldam}. This provides an efficient procedure of sampling states that would be otherwise separated by large energy barriers and with events that occur at disparate timescales. Yet, size effects are difficult to surmount. Direct atomistic simulations of phase coexistence within a single simulation cell require capturing interfaces that might occupy a significant part of the simulation. Moreover, the interface can result in significant lattice mismatch which yield strain fields with profound size-effects. Therefore, significant computational capabilities are required to include these effects correctly~\cite{yan1996,chapela1987}. 

From a modeling point-of-view, capturing the full detail of the interface region is not necessary in order to predict the overall stability in bulk phase coexistence. Indeed, various techniques have been devised which seek to determine phase boundaries without simulating interfaces directly. 
These approaches fall into two main categories: (1) a direct approach that seeks to find the free-energy of all possible phases in coexistence, and (2) an indirect approach which attempts to find phase coexistence by performing simulations in separate regions in such a way that the thermodynamic conditions of phase equilibrium are satisfied.  In the direct approach, the free energy can be calculated using thermodynamic integration to relate the fugacity via a series of simulations that connects the state of interest to a reference state with known properties\cite{mishin2003,mishin2004,williams2006,ramalingam2002,foiles1994}, although other methods to approximate the free energy such as cluster expansion method can be used\cite{sanchez1991,vanwalle2002}. This method is very reliable, but inherently inefficient as it requires a number of simulations at ``uninteresting'' state points. The difficulty of this approach lies in determining free energies (or chemical potentials) with sufficient accuracy, and as such, its implementation remains applicable to mixtures with small number of components. 

The pioneering approach by  Panagiotopoulos\cite{panagiotopoulos1992,panagiotopoulos1988,panagiotopoulos1989}, known as the Gibbs ensemble technique, proposed the use of two simulation cells for the first time. This approach
introduces particle displacements, volume fluctuations, and particle transfer between cells in such a manner that the cells are in thermal, mechanical and chemical equilibrium with each other.  This method does not require absolute free energies, rather only changes in energy that result from perturbing the system according to Monte Carlo moves. Unlike thermodynamic integration, the Gibbs ensemble technique involves only few simulations per coexistence point, hence its simplicity and efficiency  make it ideal for phase exploration. However, in its present formulation it works only in situations where particle transfer  between the cells is applicable, i.e. dilute fluids or gases, since particle insertion/deletion in solids creates point defects\cite{panagiotopoulos1992,panagiotopoulos1988,panagiotopoulos1989}. Motivated by this shortcomings, Kofke and co-workers combined direct and indirect approaches and introduced a method known as Gibbs-Duhem integration\cite{kofke1993,kofke1994}. The idea proposed by Kofke is to determine phase equilibria by integration of the Claysius-Clapeyron equation to determine phase coexistence in regions where particle insertions using the Gibbs-ensemble method fail. Various independent simulations under isobaric-isothermal conditions of each phase are performed along the saturation line. The method presumes the accurate knowledge of an initial equilibrium point at a given temperature, from this point, the pressure is adjusted to satisfy chemical potential equality according to the Gibbs-Duhem equation. This method surpasses the need to particle insertion, and has successfully explored coexistence of vapor-fluid, vapor-solid, fluid-fluid, or solid-fluid systems\cite{lamm2001,kofke1994}. 
Yet, the application to solid-solid phase equilibrium is rare, see for example \cite{Mori}. Moreover, the method requires another technique in order to find an initial starting point, quasi-harmonic approximation in solids or the Gibbs ensemble itself.

Recently, we have introduced the  Multi-Cell MC approach, abbreviated as \abb,  for simulation of phase coexistence in solids. This method was first introduced as a simulated annealing technique for energy minimization of multi component systems\cite{niu2017}. In its original form, the method imposed a strong restriction on composition variations in each cell and as such, did not allow for prediction of the phase boundaries.

To address this issue, we introduced an alternative MC move which maintains mass balance in multiple cells via application of the \textit{lever rule}. This method effectively circumvents the particle insertion/deletion moves, required by  the Gibbs ensemble MC~\cite{rao2019}. Moreover, unlike Semi-grand Canonical Simulations in the $\mu,V,T$-ensemble, which require specification of the chemical potential, \abb self-regulates the chemical potentials in each phase. However, since the balance of chemical potentials is not directly imposed, the system can get stuck in a metastable solution, which does not satisfy the condition of chemical potentials equivalence at equilibrium. In other words, application of the lever-rule alone, does not provide a well-defined criterion for when to stop the simulation.

Here, we propose an addition to the \abb method that checks for the condition of chemical potential equivalence in addition to mass balance. This condition is imposed via a predictor-corrector approach involving changes in chemical potential differences of species in different phases.


This manuscript is organized as follows. Section~\ref{sec:MCMC} introduces relevant terminology 
where we derive the thermodynamic ensemble, sampled by \abb, starting from the Gibbs ensemble and propose the additional check, via a predictor-corrector algorithm,  to confirm that thermodynamic equilibrium is reached.
Section~\ref{sec:comp} presents the details of our computational technique. Unlike our previous works~\cite{rao2019,niu2017}, we are using classical interatomic potentials to demonstrate proof of concept. Notable new features with respect to our previous work is the addition of translational MC moves to include vibrational excitations as well as volume MC moves to ensure the condition of constant pressure and temperature entirely with MC and without relying on any other method for relaxation of these degrees of freedom. 
Section~\ref{sec:results}  applies the methodology to determine phase equilibrium in model binary and multicomponent alloys. Use of interatomic potentials allows for large simulations, based on which a discussion of size effect in \abb is also provided. Finally, Section~\ref{sec:summary}  summarizes the results and  presents an outlook and potential improvements on the \abb method.

\section{Phase coexistence of mixtures}
\label{sec:MCMC}
The physical problem to be explored in this work involves finding thermodyncamic equilibrium for mixtures, specifically those phases coexisting in the solid state. At the heart of any discussion about equilibrium  phase diagrams is the Gibbs phase rule~\cite{dehoff}, which states that the number of independent intensive variables or the number of degrees of freedom of the system, F, is a function of the number of components, C and the number of phases, $\phi$:
\begin{align}
\label{eq:phase_rule}
F = C-\phi+2
\end{align}

For a unary systems in two-phase coexistence ($\phi=2,C=1$), only one independent intensive variable can be imposed. 
That is, once the temperature is specified there is only one value of pressure that will satisfy the equilibrium conditions, or vice versa.
For binary systems, on the other hand, two-phase coexistence ($\phi=2,C=2$) can be retained by imposing an additional intensive variable. 
For practical purposes, it becomes convenient to specify both pressure and temperature in order to trace out phase coexistence at various concentrations. For systems with more than one component ($C>1$), it will be convenient to search for phase coexistence under isobaric-isothermal conditions, that is the maximum number of phases that shall be considered according to the phase rule is $\phi = C$. Note that in general, the number of phases could  be larger than this, but in those cases both temperature and pressure cannot be independently specified.

\begin{figure}[ht]
  \centering
  \includegraphics[scale=.7]{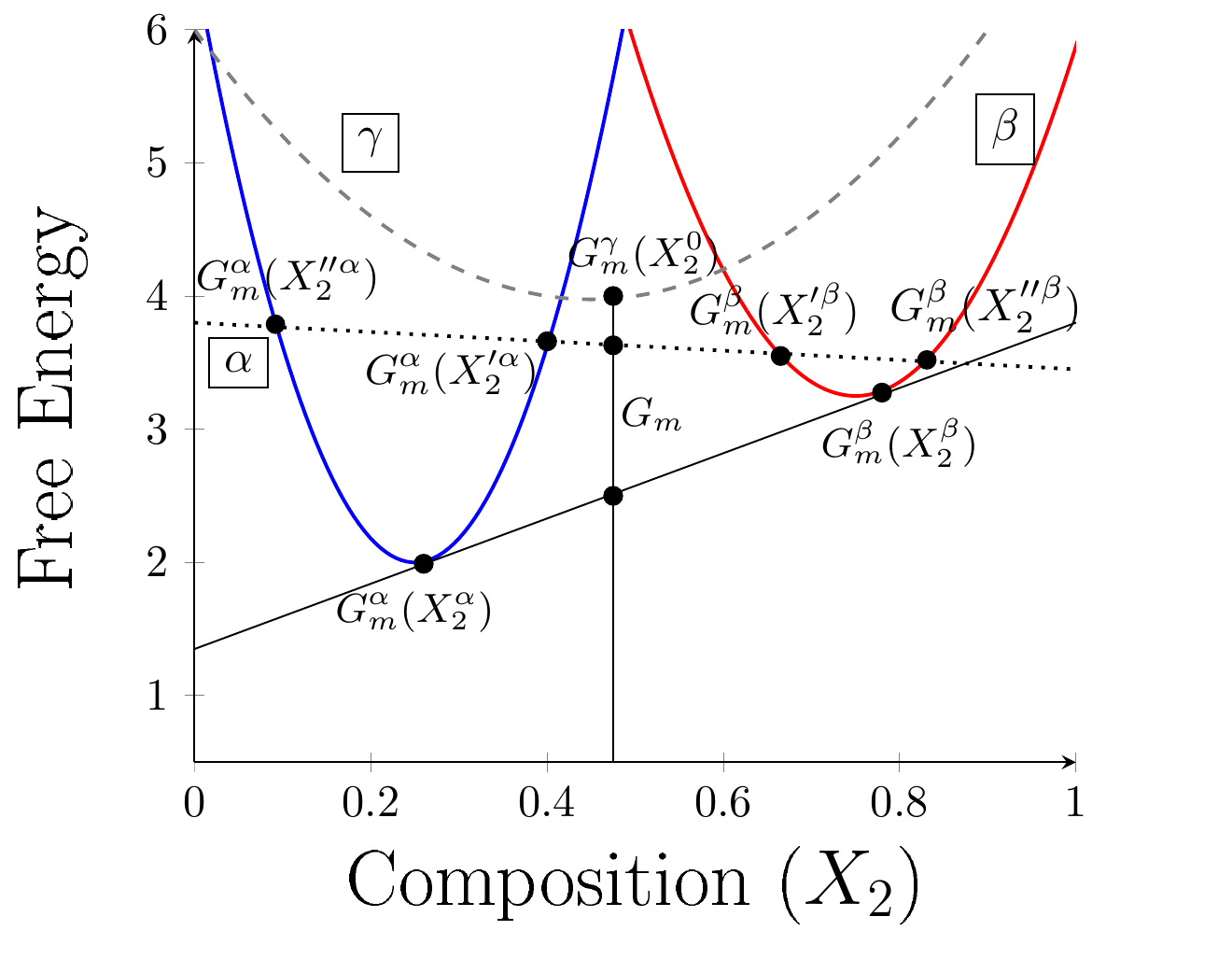}
  \caption[]{Schematic representation of free energy of mixing for a solution that separates into phases $\alpha$ and $\beta$, starting from a homogeneous phase-$\gamma$. $X_2^\alpha$ and $X_\beta^2$ are the equilibrium concentrations of phases $\alpha$ and $\beta$ respectively and are obtained from the common tangent, shown  by the (straight) solid line. The dashed line shows a non-equilibrium state where multiple concentrations can satisfy the conservation of mass. At equilibrium,  the minimum molar free energy $G_m$ lies on the common tangent where the condition of equivalence of chemical potentials in both phases is met.}
  \label{fig:diagram}
\end{figure}

Consider now a binary system that initially starts in a metastable single phase $\gamma$ whose energy minimum is given at concentration $X_2^0=n_2/N$  as shown in Fig.\ref{fig:diagram}. The notation $n_2$ describes the number of atoms of species 2 and N is the total number of particles in the system.  

This phase is unstable with respect to thermodynamic fluctuations, and the system can lower its overall energy by decomposing into neighboring phases denoted by $\alpha$ and $\beta$. In general, these phases can represent different crystal structure or amorphous phases (liquid,gas) depending on the nature of the atoms and imposed thermodynamic conditions. Conservation of the total number of atoms requires that $n^\alpha +n^\beta = N$, while conservation in the number of species implies that $n_1^\alpha +n_1^\beta = n_1$ and $n_2^\alpha +n_2^\beta = n_2$. Any mass transport between the two phases should leave the initial concentration of the entire system  unchanged. This is the basis for the \textit{lever rule}. Hence,  the overall concentration in one of the components (say component 2) can be written as

\begin{align}    
\label{eq:conc}
  X_2^o = \frac{n_2^{\alpha} +n_2^{\beta}}{N}   & = \frac{n_2^{\alpha}}{n^{\alpha}} \underbrace{\frac{n^{\alpha}}{N}}_{f^\alpha} + \frac{n_2^{\beta}}{n^{\beta}} \underbrace{\frac{n^{\beta}}{N}}_{f^\beta}  \\ \nonumber
  & = X_2^{\alpha} f^{\alpha} + X_2^{\beta} f^{\beta} 
\end{align}
where the molar fractions $f^\alpha$ and $f^\beta$ denote the relative amount of matter  in each phase, and $X_2^\alpha$ and $X_2^\beta$ denote their respective concentrations. The molar fraction plays an important role in relating extensive thermodynamic variables. 
For example, the total energy can also be written in terms of the molar free energy\cite{lupis} as
\begin{align}
\label{eq:Gm}
  G_m  & = \frac{G^{\alpha}}{n^{\alpha}} \frac{n^{\alpha}}{N}  +\frac{G^{\beta}}{n^{\beta}} \frac{n^{\beta}}{N} \\
   & = G_m^\alpha ~ f^{\alpha} ~+~ G_m^\beta ~ f^{\beta}
\end{align}
where $G_m^\alpha$ and $G_m^\beta$  are the molar free energy of phase $\alpha$ and $\beta$ respectively. Graphically, the value of the molar energy will lie along the  tie-line connecting the energy between the two phases, and its position along the line is constrained by the initial concentration of the overall system. For the scenario depicted in Fig.\ref{fig:diagram}, a lower energy can be obtained if the system spontaneously decomposes into two phases $\alpha$ and $\beta$. Two possible states for arbitrary concentrations are shown with lines connecting the corresponding energy values in each phase. The values connected by the dashed line are not in equilibrium since the gradients in the energy of the phases with respect to concentration, i.e chemical potentials, are not equal to one another. On the other hand, points connected by the solid line can be considered to be in thermodynamic equilibrium according to the common tangent criterion. Once the overall system attains a minimum in free energy it can be shown that that the subsystems will be in thermal, mechanical, and chemical equilibrium with one another\cite{dehoff}. The \abb  seeks to satisfy these conditions. 
Before describing this method, it is useful to first understand the main ideas behind the Gibbs Ensemble, upon which the \abb formalism is rooted.

\subsection{Gibbs Ensemble Monte-Carlo}
The  idea behind the Gibbs Ensemble Monte Carlo introduced by Panagiotopoulos\cite{panagiotopoulos1988,panagiotopoulos1989,panagiotopoulos1992,Panagiotopoulos1995}
was to model phase coexistence by imposing conditions of thermal equilibrium in separate cells representing different phases. 
The equillibrium phase diagram of mixtures is typically measured under constant temperature and pressure. Therefore, it is appropriate to perfrom the simulations under the isobaric-isothermal condition, where the volume of the cells are allowed to change independently in order to reach the overall target pressure. For a full a review of the Gibbs ensemble approach, the reader is referred to \cite{panagiotopoulos1988,panagiotopoulos1992,Panagiotopoulos1995,panagiotopoulos1989}

We begin with the partition function for the isobaric-isothermal Gibbs Ensemble for a mixtures in two cells, as given in~\cite{green}. The partition function $Q^{Gibbs}_{NPT}$ provides the number of ways of arranging $N=n_1 + n_2$ particles amongst the two regions (cells) $\alpha$ and $\beta$ while $P$ and $T$ are kept constant: 

  \begin{align}
    \label{eq:GNPT}
    Q^{Gibbs}_{NPT} = & \frac{1}{n_1! \Lambda^{3n_1} V_o} \frac{1}{n_2! \Lambda^{3n_2} V_o} \sum_{n_1^\alpha=0}^{n_1} 
    \frac{n_1!}{n_1^\alpha! n_1^\beta!} \sum_{n_2^\alpha=0}^{n_2} \frac{n_2!}{n_2^\alpha! n_2^\beta!} \int_0^\infty dV^\alpha \exp\Big( \frac{-P V^\alpha}{kT} \Big)\int_0^\infty dV^\beta \exp\Big( \frac{-P V^\beta}{kT} \Big) \nonumber \\
    & \times \int d (r^{\alpha}_1)^{n_1^\alpha} \int d (r^{\alpha}_2)^{n_2^\alpha} 
    \int d (r^{\beta}_1)^{n_1^\beta} \int d  (r^{\beta}_2)^{n_2^\beta}~\exp\Big[ \frac{-U^\alpha(n^\alpha)}{kT} \Big] ~\exp\Big[ \frac{-U^\beta(n^\beta)}{kT} \Big]
  \end{align}
  where $\Lambda$ is  the thermal de Broglie wavelength, $V_o$ is a unit of volume chosen to make the partition function dimensionless~\cite{wood},  $k_B$ is the Boltzmann constant,  $r_i^\nu$ are the positions of particles of type i in subsystem $\nu$, and $U^\nu(n^\nu)$ is the energy of subsystem $\nu$.  It is convenient to use rescaled coordinates $ \xi_i^{\nu} = r_i^{\nu}/L^\nu $ to write volume and energy terms independently.  
The ensemble average of a variable $A$ can be written using the above partition function as:

  \begin{align}
    \label{eq:aGNPT}
    \langle A \rangle^{Gibbs}_{NPT} = & \frac{1}{Q^{Gibbs}_{NPT}}\frac{1}{n_1! \Lambda^{3n_1} V_o} \frac{1}{n_2! \Lambda^{3n_2} V_o} \sum_{n_1^\alpha}^{n_1} 
    \frac{n_1!}{n_1^\alpha! n_1^\beta!} \sum_{n_2^\alpha}^{n_2} \frac{n_2!}{n_2^\alpha! n_2^\beta!} \int_0^\infty dV^\alpha \exp\Big( \frac{-P V^\alpha}{kT} \Big) (V^\alpha)^{n^\alpha} \nonumber \\
    & \times \int_0^\infty dV^\beta \exp\Big( \frac{-P V^\beta}{kT} \Big) (V^\beta)^{n^\beta}   \nonumber \\
   & \times  \int d (\xi^{\alpha}_1)^{n_1^\alpha} \int d (\xi^{\alpha}_2)^{n_2^\alpha}  
    \int d (\xi^{\beta}_1)^{n_1^\beta} \int d  (\xi^{\beta}_2)^{n_2^\beta}~\exp\Big[ \frac{-U^\alpha(n^\alpha)}{kT} \Big]~\exp\Big[ \frac{-U^\beta(n^\beta)}{kT} \Big] A 
  \end{align}


,whence it follows that a probability density for this ensemble can then be written as~\cite{green}:  
\begin{align}
\label{eq:psdB}
  \wp^{Gibbs}_{NPT} =& \exp\Big[ 
    \ln \Big( \frac{n_1!}{n_1^\alpha! n_1^\beta!} \Big) +    \ln \Big( \frac{n_2!}{n_2^\alpha! n_2^\beta!} \Big) 
+   n^{\alpha} \ln V^\alpha \nonumber \\ 
&+   n^{\beta} \ln V^\beta - \frac{P V^\alpha}{kT} - \frac{P V^\beta}{kT} - \frac{U^\alpha}{kT} - \frac{U^\beta}{kT}
\Big] 
\end{align}

In accordance with the Metropolis criteria\cite{metropolis}, the probability of accepting a new configuration in this ensemble is given by  $\displaystyle min\{1,\frac{\wp'}{\wp}\}$, where $\wp$ and $\wp'$ are the probability densities of the new and old configurations respectively. Therefore, the corresponding  acceptance criteria for different MC moves can be derived as follows.

The simplest type of MC moves, under the above ensemble, involves  translational displacement of the atoms.  Within a given cell, say cell $\nu$, a translational perturbation of the atoms results in a change in the potential energy ($\Delta U^\nu$). An acceptance probability for this type of move is given by
\begin{align}
\label{eq:transl}
 \wp^{acc}_{transl.} =  min\{1, e^{ -(\Delta U^\nu)/k_B T} \},
\end{align}
 
Another move with a similar acceptance criteria involves sampling the configurations of the atomic species inside the same cell. This can be achieved by swapping the species on any two atoms within a given cell; we shall refer to these moves as {\it intra-cell} swap. The rearrangement of atomic species (for a pair of distinct atoms) results in a change in the potential energy of the cell $\nu$. Thus, an acceptance probability similar to the one above is given by  
\begin{align}
\label{eq:intra}
 \wp^{acc}_{intra-cell} =  min\{1, e^{ -(\Delta U^\nu/k_B T} \},
\end{align}
where $\Delta U^\nu$ is the change i potential energy, this time, due to the inter-cell swap.

Next, volume changes are performed on the two cells to arrive at a target pressure on the whole system. 
In this case, volume rearrangements of the two regions $V^{\alpha'} \rightarrow V^\alpha + \Delta V^\alpha$ and $V^{\beta'} \rightarrow V^\beta + \Delta V^\beta$ can be performed on each cell, where volume changes $\Delta V^\alpha$ and $\Delta V^\beta$ are typically chosen from a uniform distribution between a minimum and maximum value. Thus, the acceptance probability for fluctuating volumes is given by
\begin{align}
\label{eq:vol}
\wp^{acc}_{volume} =  min\{1,& \tiny{(\frac{V^\alpha + \Delta V^\alpha}{V^\alpha})^{n^\alpha} (\frac{V^\beta + \Delta V^\beta}{V^\beta})^{n^\beta}} \nonumber \\ 
& e^{-(\Delta U^\alpha + \Delta U^\beta + P (\Delta V^\alpha + \Delta V^\beta)/kT)} \}
\end{align}

Notice that in the moves described above, the acceptance criteria is identical to the product of two independent NPT-ensembles corresponding to two uncorrelated systems. This is no longer the case when a MC move couples the two systems, 
for example  during direct mass transfer from one cell to the other. Suppose that a particle of species 1 from phase $\alpha$ is transferred to phase $\beta$, i.e. $n^{\alpha'} \rightarrow  n^{\alpha} -1$ and $n^{\beta'} \rightarrow  n^{\beta} +1$, while $n^{\alpha~'}_1 \rightarrow  n^{\alpha}_1 -1$, and $n^{\beta~'}_1 \rightarrow  n^{\beta}_1 +1$.  
From the ratio of probabilities given by Eq.{\ref{eq:psdB}  the acceptance criterion for this move can be written as,
\begin{align}
\label{eq:transfer}
\wp^{acc}_{transfer} =  min\{1, \frac{n_1^\alpha V^\beta}{(n_1^\beta+1) V^\alpha}e^{ -(\Delta U^\alpha + \Delta U^\beta)/k_B T} \}
\end{align}

Vapor-liquid and liquid-liquid coexistence of mixtures have been successfully modeled using the above ensemble\cite{wood,panagiotopoulos1989}. Yet this technique has been shown to be inefficient as the density increases since the insertion of particles of particle for systems of considerable densities (i.e. solids) these moves are mostly rejected.

However, in the case of mixtures, transferring particles between the two cells can be done indirectly. A particle identity exchange approach occurs by changing a particle of type 1 into one of type 2 in one of the two regions with a simultaneous reverse change in the other cell ~\cite{Panagiotopoulos1995}.  This inter-cell swap, or exchange move is accepted with a probability
\begin{align}
\wp^{acc}_{exchange} & = min\{1,\frac{n_1^\beta n_2^\alpha}{(n_1^\alpha+1) (n_2^\beta +1)} e^{ -(\Delta U^\alpha + \Delta U^\beta)/k_B T} \}
\label{eq:inter}
\end{align}
Recently, Niu {\it et al}~\cite{niu2017} have used a similar MC move in a multi-cell MC relaxation of solids. 
The \textit{exchange} MC move has the advantage of  maintaining  equal  chemical potentials between the cells, but poses an unnecessary additional restriction on the final equilibrium compositions.
Assuming the overall chemistry of the whole system ($\alpha + \beta$) to be constant, conservation of mass requires that the total number of each species among all cells be constant, i.e $n_1^\alpha+n_1^\beta=n_1$ and $n_2^\alpha+n_2^\beta=n_2$ (See Section.\ref{sec:MCMC}). However, $n_1^\alpha$ and $n_1^\beta$ ( and similarly $n_2^\alpha$ and 
$n_2^\beta$) can change independently as long as their total sum is conserved.
On the other hand, during the \textit{exchange} move, an increase (decrease) in $n_1^\alpha$ is always accompanied by a similar decrease (increase) in $n_1^\beta$ which restricts the compositional variation in each cell. 
In reality, both the composition and size of each phase should be able to change independently as to find most stable phase.
In order to satisfy the conservation of mass, the relative amounts of each phase can change and the corresponding phase fractions (relative sizes) can be obtained from the lever rule.

Note that the insertion/deletion move of the original Gibbs ensemble solves this problem. However, insertion/deletion of atoms in crystalline systems creates vacancies and interstitials, and makes it impossible to apply the original Gibbs ensemble MC to crystalline solids. To overcome this restriction, Niu and coworkers\cite{rao2019} proposed a new method to explore variable compositions and the corresponding phase fractions. We briefly describe this method next and show its relationship to the original Gibbs ensemble.

\subsection{Multi-Cell Monte Carlo}
In the \abb method for phase prediction, Niu and coworkers\cite{rao2019} have recently proposed a novel 
method to eliminate the composition restriction, discussed above. Instead of exchanging particles between different cells, the composition of each cell is randomly changed, while the overall composition of the system is maintained numerically by enforcing the lever rule over the different phases. This way, the atomic fractions within each cell are allowed to fluctuate independently to the extent that they don't violate the conservation of mass across all cells/phases.

In a multicomponent system, the mass balance qquation can be written by generalizing Equation~\ref{eq:conc}, as
\begin{align}    
\label{eq:masseqn}    
  X_1^o & = X_1^{\alpha} f^{\alpha} + X_1^{\beta} f^{\beta} + \cdots \nonumber \\
  X_2^o & = X_2^{\alpha} f^{\alpha} + X_2^{\beta} f^{\beta} + \cdots \nonumber \\
  X_m^o & = X_m^{\alpha} f^{\alpha} + X_m^{\beta} f^{\beta} + \cdots 
\end{align}
 where the indices $\nu = \{\alpha,\beta,\cdots,\phi\}$ represent phases, indices $i= \{1,2,\cdots,m\}$ denote chemical components, the terms $X_i^o=(n^\alpha_i+n^\beta_i+...)/N$ are the initial concentration of the $i^{th}$ species for the overall system where 
N is the total number of particle in the system, viz $N = (n^\alpha+n^\beta+...)$, and the extensive variable $f^\nu = n^\nu/N $ (molar fraction) captures the relative amount of phase $\nu$. Niu {\it et al}  proposed to solve for this scaling variable numerically, by writing Equation (\ref{eq:masseqn}) in matrix form and solving it via Cramer's rule\cite{rao2019}. Obtaining the phase fractions throughout the simulation, provides a way to mimic the growth of new phases, without performing mass transfer physically.

Mimicking mass transfer has been done in the past.  Allen {\it et al} used the concept of a ``pseudo-Gibbs ensemble'' to simulate mass transfer in dense phases (liquid-vapor) by altering the volumes of two cells in such a manner that the response is equivalent to particle transfer\cite{allen}. In a similar vein, the molar fractions are determined self-consistently in \abb by changing the chemical composition of the various phases (cells) provided they satisfy the lever-rule constraint (Eq. \ref{eq:masseqn}). In this manner, the evolution of the molar fractions captures the relative growth of the phases, with respect to their initial values. 

Composition fluctuations within each cell is achieved via a \textit{flip} move~\cite{rao2019}. During a \textit{flip} move, a particle is randomly selected and its chemical identity is changed to another type, while enforcing the total composition of the system via the application of the lever rule. The \textit{flip} moves are similar to those implemented in the Semi-Grand Canonical Ensemble (SGCE)~\cite{frenkel} where atomic species are swapped with an imaginary reservoir, while keeping a chemical potential gradient between the system and the reservoir fixed. In the SGCE, the chemical composition is adjusted by choosing the appropriate chemical potentials, often done by trial and error, which can be time consuming and not very convenient when a specific composition is targeted. In the \abb approach, however, the other cells are used to accommodate the mass transfer that results from the virtual mass exchanges, and an initial concentration can be set in advance where phase coexistence is to be investigated.

In order, to write acceptance rules based on this approach, all extensive variables are re-written in the probability density of the Gibbs-NPT ensemble (Eq.\ref{eq:psdB}) to show the explicit dependence of molar fraction variables. For clarity, let us focus on two phases ($\alpha$ and $\beta$). The total number of atoms in phase $\alpha$ is given by $n^\alpha = f^\alpha N$, while the number of component 1 can be written as  $n_1^\alpha = N f_\alpha X_1^\alpha$, where $X_1^\alpha = n_1^\alpha/n^\alpha$. Similarly, the potential energy and volume of phase $\alpha$ can be written as  $U^\alpha = N f_\alpha u_\alpha$, and $V^\alpha = N f_\alpha v^\alpha$ respectively, where $u^\alpha= U^\alpha/n^\alpha$ and $v^\alpha= V^\alpha/n^\alpha$. Using similar expression for phase-$\beta$, Eq.\ref{eq:psdB} can be written as

\begin{widetext}
\begin{align}
\label{eq:psdB2}
  \wp^{Gibbs}_{NPT} = \exp\Big[ & \ln \Big( \frac{(NX^0_1)!}{(N f_\alpha X_1^\alpha)! (N f_\beta X_1^\beta)!}\Big) +  \ln \Big( \frac{(NX^0_2)!}{(N f_\alpha X_2^\alpha)! (N f_\beta X_2^\beta)!}\Big) \nonumber \\ 
&    +   f^\alpha N \ln (N f_\alpha v^\alpha) +   f^\beta N \ln (N f_\beta v^\beta) - N [ P (f^\alpha v^\alpha + f^\beta v^\beta)+f^\alpha u^\alpha+ f^\beta u^\beta]/kT
\Big] 
\end{align}

\end{widetext}
The above expression can be simplified by making use of the Stirling approximation:  \textbf{$\ln N! \approx N \ln N - N$} to reduce the
above expression to 

\begin{eqnarray}
\label{eq:psdB3}
  \wp(f^\alpha,N,P,T)  && \sim e^{-N[f^\alpha u^\alpha+ f^\beta u^\beta + P (f^\alpha v^\alpha + f^\beta v^\beta)]/(k_B T)} \nonumber \\
  && \times ~ e^{ -N f^\alpha [X_1^\alpha \ln X_1^\alpha + X_2^\alpha \ln X_2^\alpha-\ln v^\alpha]}  \nonumber \\
  && \times ~ e^{ -N f^\beta [X_1^\beta \ln X_1^\beta + X_2^\beta \ln X_2^\beta-\ln v^\beta]}
\end{eqnarray}
A few constant terms where the molar fraction does not appear explicitly have been omitted for clarity. Note that the Stirling formula above is valid for large $N$ values. For small $N$ values, we recommend including the next order terms, or using alternative approximations for increased accuracy.


The acceptance criterion for a flip move can now be defined according to the above pseudo-Boltzmann factor as follows.
Once a particle (or a group of particles ) is (are) randomly flipped, new values of  $f'^\alpha$ and $f'^\beta$ can be calculated from Eq. \ref{eq:masseqn}, where the initial concentration ($X^0_i$ ) is kept fixed, and the new molar energies $u'^\alpha$ and $u'^\beta$ and molar volumes $v'^\alpha$ and $v'^\beta$ are obtained from the simulation. The flip acceptance criterion is given by the ratio of the statistical weights of states before and after flipping the particles, viz
\begin{align}
\label{eq:flip}
 \wp^{acc}_{flip}   =  &  min\{1,\frac{\wp(f'^\alpha,N,P,T)}{\wp(f^\alpha,N,P,T)}\}  \\ \nonumber 
 = &   min\{1,e^{-\Delta G_m/k_B T} \}
\end{align}
where the second line above has been written as a Boltzmann probability with 

\begin{align}
\label{eq:metro}
\Delta G_m  =   & N\sum_\nu  (f'^\nu u'^\nu - f^\nu u^\nu) + N\cdot P \sum_\nu(f'^\nu v'_\nu - f^\nu v^\nu) \nonumber \\
& - N k_B T \Big[ \sum_\nu (f'^\nu \ln v'_\nu -f^\nu \ln v^\nu)  \nonumber \\ 
& + \sum_\nu f'^\nu \sum_j  X'^\nu_j \ln X'^\nu_j - \sum_\nu f^\nu \sum_j  X^\nu_j \ln X^\nu_j  \Big].
\end{align}
The primed (unprimed) quantities above refer to quantities after (before) the flip move.

\textit{Deriving the most general acceptance criterion for \abb is the first new result of this paper.} Using this acceptance criterion, the Metropolis MC can search for a minimum in the molar free energy, provided that a self-consistent value for the molar fractions can be obtained from the lever-rule constraint. In this manner, phase growth is mimicked by coupling the phases in their molar energy which provides a great leap forward towards finding solutions to solid phases self-consistently, in the spirit of the Gibbs ensemble approach, without having to resort to determination of the free energy directly.  

However, in practice, relying on mass balance only, does not define a proper stopping condition to ensure the equilibrium solution is found. Recall Figure.\ref{fig:diagram}. The flip moves in \abb are equivalent to spanning the concentration space along each parabola. Let's say flip moves have brought the concentration of species 2 in phase (cell) $\alpha$ to $X_2^{\prime \alpha}$ and that in phase $\beta$ to $X_2^{\prime \beta}$. The total molar energy will then lie along the  tie-line connecting the energy between the two phases (the dashed line), and its position along the line is constrained by the initial concentration of the overall system, i.e $X_2^0$. It immediately follows that, $X_2^{\prime \prime \alpha}$ and $X_2^{\prime \prime \beta}$ will result in exactly the same molar energy.

In fact, except for the common tangent, any  line between the two phases, intersects each parabola at two points. In other words, for each molar energy $G_m$ the mass balance equation does not have a unique solution. As $G_m$ is minimized, the dashed line approaches the common tangent, with a unique solution.  \abb seeks to minimize $G_m$ numerically, by spanning different points along each parabola. During this process, very different sets of concentrations and molar fractions can give exactly the same energy value. Consequently, even though the energy may seem to have reached a plateau the molar fractions may not have converged to the equilibrium values. This issue becomes particularly important when $G_m$ has decreased, within numerical accuracy, close to the minimum value, but is not exactly at the minimum and still has degenerate sets of concentrations and molar fractions. Therefore, it is necessary to define a stopping criterion for the simulation to ensure that in the end, the solution closest to the common tangent condition has been found. The common tangent condition is equivalent to the balance of chemical potentials at equilibrium. 

Note that this issue does not arise in Gibbs ensemble, since there, mass is physically transferred between the cells at any given step, so that there is no ambiguity about the composition and size of each phase. However, in \abb mass transfer is only mimicked through solving the mass balance equation, which can have more than one solution except exactly at equilibrium.

Therefore, to avoid getting trapped into these solutions and to define a proper stopping criterion for \abb, here we propose a consistency check on the energy variations as a result of the virtual mass changes. This check is performed  via a predictor-corrector approach that aims to penalize solutions not satisfying the common-tangent criterion as follows. 
Every time a number of particles are flipped ($i \rightarrow j$), a predicted energy change is compared with the actual energy change in a modified acceptance criteria given by (see appendix \ref{app:b}):
\begin{eqnarray}
  \label{eq:Gw}
  \Delta \tilde{G}_m  && = \Delta G_m \cdot (1-w) \nonumber \\ 
  && + ~ w\cdot\frac{\delta n_{ji}^\alpha}{2} [f^\alpha (\Delta \mu_{ij}^\alpha -\Delta \mu_{ij}^\beta)  + f'^{\alpha} (\Delta \mu'^\alpha_{ij}-\Delta \mu'^\beta_{ij})] \nonumber \\
\end{eqnarray}
where $w$ is a numerical weight between 0 and 1,  $\delta n_{ij}$ is the number of replacements ($ i\rightarrow j $),
and $\Delta \mu_{ij} \equiv \mu_i - \mu_j$ is the {\it difference} in the chemical potential between species i and j.  
Note that in the limit of $w\rightarrow0$, the acceptance criterion reduces to that of Eq.\ref{eq:metro} and the predicted-corrected terms are not used.
On the other hand, for \textit{non-zero} values of the weights, the accepted trial moves will now attempt to minimize the molar energy difference $G'_m-G_m$, and the difference in the {\it chemical potential gradients} between the phases, ie  $(\Delta \mu_{ij}^\alpha -\Delta \mu_{ij}^\beta)$ .
We emphasize that the equality of the chemical potential \textit{differences}, is a weaker form for the equilibrium condition than the equality of the chemical potentials. In other words, it is a necessary but not sufficient condition for equilibrium.  

In order to satisfy the common tangent criterion within \abb,   $\Delta \mu_{ij}$ for each phase needs to be evaluated with a high degree of accuracy. To do so, a variation of Widom's test particle proposed by Frenkel  \cite{frenkel,frenkel1987}) method is employed (see Appendix\ref{app:a}). One major drawback of this method is that it requires large statistics to determine the chemical potential gradients accurately and this can be computationally expensive. Nevertheless, once the system reaches equilibrium, it can be shown that changes  in the chemical potential caused by fluctuations in chemistry will tend to be centered around a mean value ($\Delta \bar{\mu}_{ij}$), with fluctuations around that mean given by ($\sigma_{\Delta \mu_{ij}}$).
For this reason, it is not necessary to evaluate the chemical potential (differences) every time a flip is made, rather a random sampling of the chemical potential on each cell is enough to determine whether the system is approaching thermodynamic equilibrium, and if not add an energy penalty to steer the search towards satisfying the chemical potential constraint. \textit{Defining a stopping criterion for \abb via this predictor-corrector approach is the second major result of this paper.}

In the following sections, we demonstrate that the modified algorithm introduced here can avoid falling into metastable solutions during \abb, by seeking to find a minimum in free energy and common tangent of the two ( or more) phases simultaneously.  The computational scheme used to perform these simulations is described next.

\section{Computational Scheme}
\label{sec:comp}
All simulations are performed using the LAMMPS\cite{plimpton1993} package with in-house  routines that take advantage of the partition framework to run  multiple cells in parallel. Fig  \ref{fig:flow} shows a schematic flow chart of various steps employed in this work. Prior to starting the Monte Carlo search, an initial concentration of the cells is set by the user, and the cells are brought to the target pressure and temperature via Nose-Hoover NPT ensemble for roughly 50 ps using standard couplings for the thermostat and barostat. This step is not necessary, but is a convenient starting point to for the \abb algorithm described next. 

Thermal equilibrium is introduced by using either MC translational moves or molecular dynamics (MD) coupled to a thermostat/barostat. In the first case, a random displacement vector is chosen on a given particle ($\vec{r}_{max} = 0.20$ \AA) and the moves are accepted according to the MC acceptance (Eq.\ref{eq:transl}).
Efficient sampling using this approach requires $O(N)$ evaluations, where $N$ is the number of atoms in the cell. However, a less expensive approach can be used to introduce the translational movements via molecular dynamics while temperature of the system is controlled by a thermostat.  For this work, we chose a Langevin thermostat with a coupling timescale of  $\tau=100 \times \delta t$ , where $\delta t = 0.001$ ps is the MD timestep, and the system is evolved for 10 $\times \tau$ to reach thermal equilibrium.

\begin{figure}[ht]
  \centering
  \includegraphics[scale=0.70]{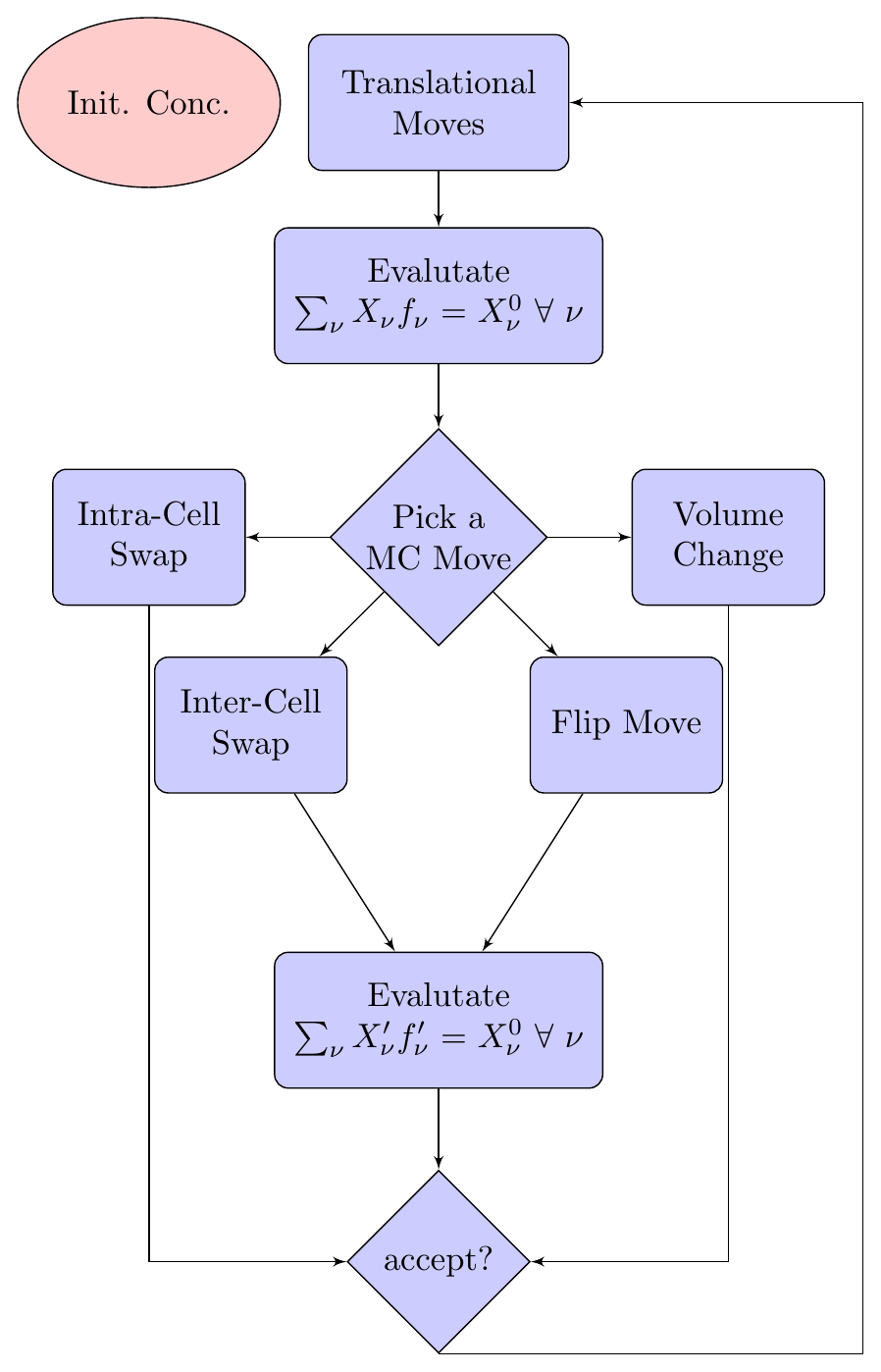}
  \caption[]{ Schematic flow chart of the \abb algorithm}
  \label{fig:flow}
\end{figure}

After the translation moves are sampled (either by MC scheme or MD using a thermostat), a random number is used to pick between the following MC moves:

\begin{enumerate}
\item Intra-cell swap (Eq.\ref{eq:intra})
\item Inter-cell swap, or exchange (Eq.\ref{eq:inter})
\item Flip move  (Eq. \ref{eq:flip})
\item Volume change  (Eq.\ref{eq:vol}). Only isotropic volume changes are perfromed, but in principle  shape change moves are possible as well. In that case the acceptance criterion should consider the full stress tensor rather its trace (i.e pressure). Moreover, volume changes and particle velocities can also be controlled via (reversible or stochastic) MD thermostat and/or barostats schemes. We reproduced most of the results with hybrid MD/MC except for the cases of small cells at higher temperature where MD results in  premature melting. This is discussed in Section~\ref{sec:results}. 

\end{enumerate}
The first three moves in the above list are selected with roughly equal probability, whereas the volume changes is only selected 10\% of the time. After a MC move in the above list is attempted, the system is again evolved with the translational moves as before; one complete loop in Fig.\ref{fig:flow} will be referred to as a {\it cycle}. Note, that in addition to the Metropolis acceptance criterion, MC moves involving mass transfer between cells are accepted \textit{only if} the molar fractions yield a physical solution after the proposed moved. This is because it is possible to obtain a numerical solution outside the range: $0 \leq f^\nu \leq 1 $, depending on the starting concentration.  The computational scheme is now applied to various alloys systems using the embedded atomic methods (EAM) to describe chemical bonding of various metallic systems as described next. Details of the EAM potential and size of various simulations  are presented in the corresponding sections.

\section{Results}
\label{sec:results}
We apply the generalized \abb method to compute phase boundaries of a few binary alloys with a miscibility gap and study the effect of simulation size on the predictions. We then apply the method  to a quaternary model system as a proof of concept. The results are compared against thermodynamic integration and/or experimental data when available.

\subsection{Prediction of Au-Pt alloy phase boundary using \abb}
In this section, we reproduce the phase diagram of a model binary (Au-Pt) system showing a miscibility gap with a Pt-rich FCC solid solution phase and an Ag-rich FCC solid solution phase. 

The computational scheme described in the previous section is now applied at various target temperatures to arrive at a solid-solid phase coexistence under zero pressure, using two simulation cells. Here, the interatomic potential developed by O'Brien \textit{et al} is employed, where the energies of the interatomic potential were parameterized using force matching with density functional theory on inter-metallics and disordered configurations\cite{obrien2018}. 

Fig.\ref{fig:vars} shows the evolution of various thermodynamic variables using 108 atoms per cell for two phases.
In addition, this figure examines the role of weight $w$ used in the predictor-corrector scheme using two values of $w=0$ vs $w=0.75$. The chemical potential gradients were randomly evaluated with a frequency of 3\% of the simulation duration. 

\begin{figure}[ht]
  \centering
  \includegraphics[scale=0.40]{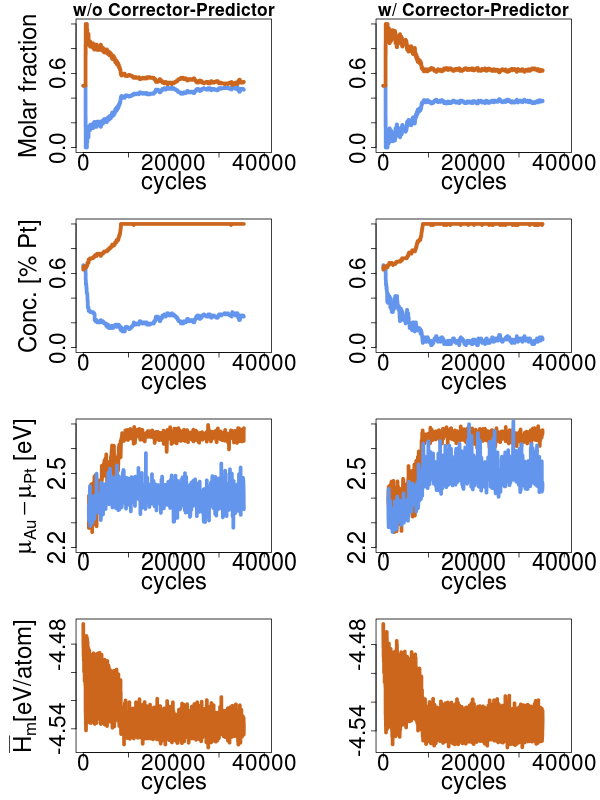}
  \caption[]{ 
    Comparison of various thermodynamic quantities vs the number of cycles for a Au-Pt system (T=400 K, P=0 GPa) \textit{with} and \textit{without} the predictor-corrector approach. Starting from top to bottom, molar fraction, concentration, chemical potential difference, and molar enthalpy $\bar{H}_m$.
    Note that the two colors correspond to distinct phases
    }
    \label{fig:vars}
\end{figure}

Comparing the evolution of the two cells, we see that the corresponding concentration and  molar fractions grow initially in similar directions, but after about 10,000 cycles an important bifurcation occurs. 
In the case where the predictor-corrector correction is applied, the Au-rich phase continues to grow after reaching 10,000 cycles but the molar fractions appear to remain fixed hereafter. 
Conversely, in the case where the predictor-corrector is not used, the molar fractions continue to evolve continuously until they get closer to one-another, while the concentration in the Au-rich 
phase  remains around the $X_{Pt} \sim 20 \%$ mark.

An estimate to the free energy of the coupled system can be obtained from an ensemble average of the enthalpies, i.e. $\bar{H}_m  = \langle f^\alpha h^\alpha_m +f^\beta h^\beta_m \rangle $ where $h^\alpha$ and $h^\beta$ are the enthalpies (per atom) of the phase $\alpha$ and $\beta$ respectively (see Eq. \ref{eq:aGNPT}). Note that an absolute value of the free energy requires defining an integration path that includes a thermodynamical state of well-defined free energy  \cite{frenkel}. Nevertheless, the ensemble average of $\bar{H}_m$ can be used to estimate the converged molar free energy up to a constant of integration.

As shown at the bottom of Fig.\ref{fig:vars}, the two solutions attempt to lower the overall molar enthalpies of the binary system. In fact, the final plateau in the molar enthalpies are essentially identical in both cases. Yet, in the case where the predictor-corrector is {\it not used}, the final solution obtained cannot be considered to be in equilibrium since the chemical potential gradients show a larger difference, whereas introducing an internal consistency check on the common tangent (i.e. $w \ne 0$) leads to  a different state with the same overall enthalpy, but with closer chemical potential differences.  

Fig.\ref{fig:wPtAu}(a) traces the coexistence of the two phases using various weights in \abb (symbols), compared against the values obtained by the standard method of thermodynamic integration by O'Brien \textit{et al.}~\cite{obrien2018} (dashed lines) and experimental data (solid lines)\cite{vesnin1988}. The final concentration of each cell is obtained by averaging over the last $20 \%$ of the simulation cycles and the error bars illustrate the standard errors of the concentration values.

Without the predictor-corrector correction, i.e. 
$w=0$, the reproduced phase diagram matches the reference curves only in some cases. An optimal value of $w \sim  0.75$ emerges from considering different weights, where the predictions (squares) reproduce entirely the main features of the phase diagram compared to the references. Therefore,  while it is possible that the algorithm without the predictor-corrector  finds the equilibrium state, it is necessary to use the predictor-corrector approach  to improve the search efficiency and screen the solution closest to equilibrium. 

Fig.\ref{fig:wPtAu}(b) shows the estimates of the molar enthalpy $\bar{H}_m$ vs temperature for various weight values, averaged over the last 20\% of the simulation cycles.  It is evident that the optimal weight of ($w \sim  0.75$ ) also yields the lowest values of this quantity over the entire temperatures range explored. 

\begin{figure}[t]
  \centering
  \includegraphics[scale=0.220]{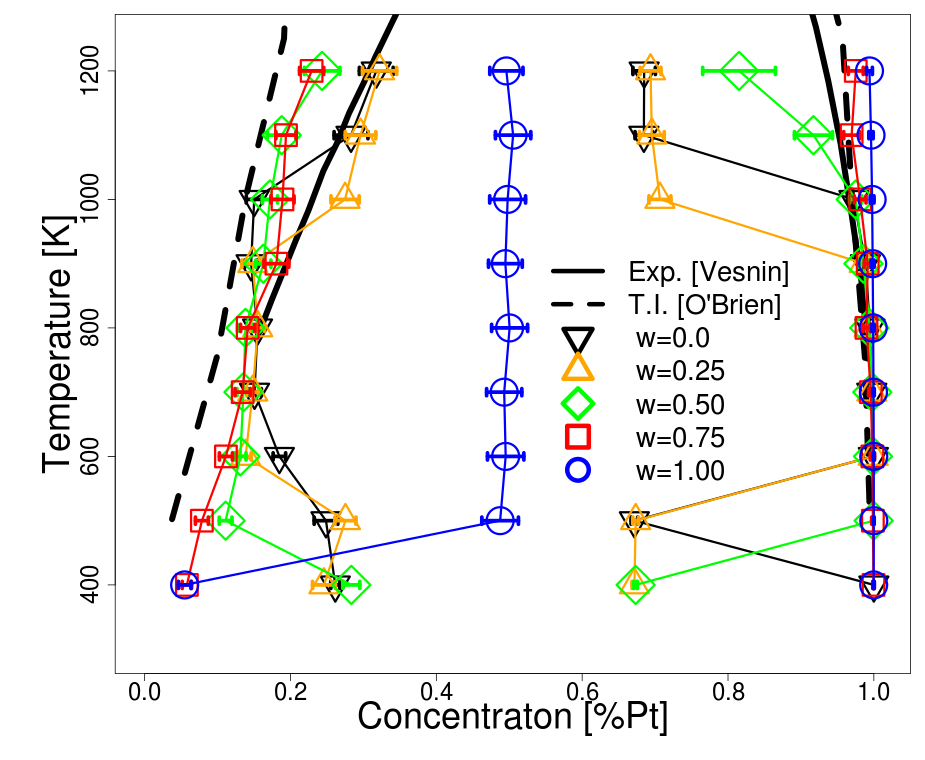} \\  \hspace{1cm}{\Large \textbf{(a)} } \\
  \includegraphics[scale=0.220]{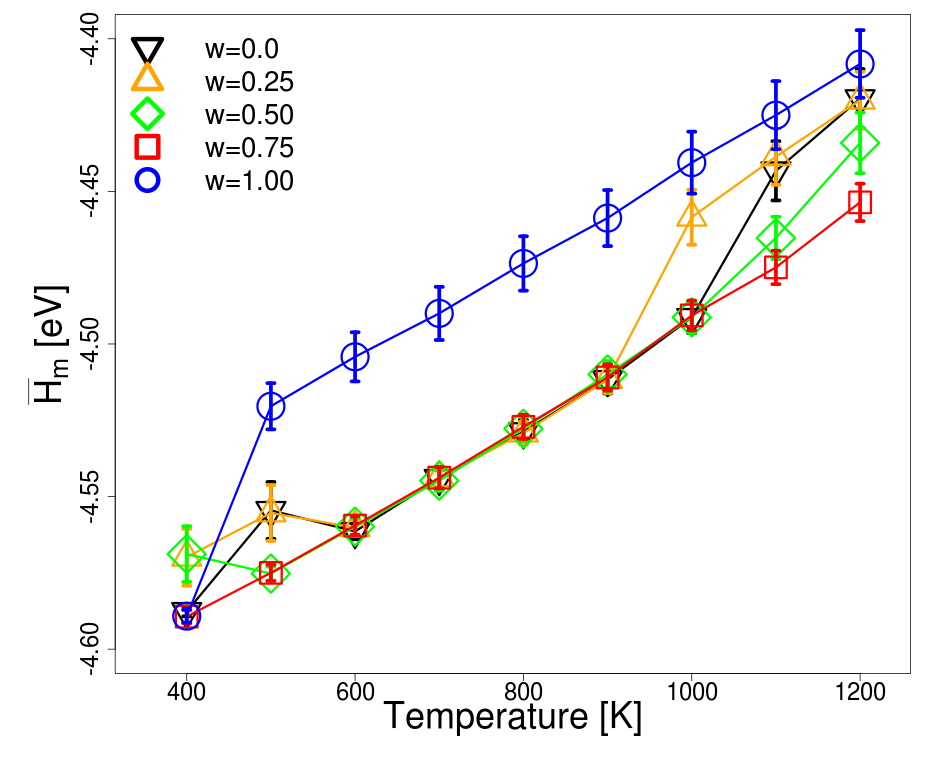} \\ \hspace{1cm}{\Large \textbf{(b)} } 
  \caption{ Effect of the predictor-corrector scheme on \abb predictions.
    (a) shows the calculated phase diagram of the Au-Pt system (symbols) for different weights $w$ used in the predictor-corrector approach. Predictions are compared against Thermodynamic Integration calculations of O'Brian \textit{et al.}~\cite{obrien2018}(dashed-lines) and experimental data from Vesnin \textit{et at.}~\cite{vesnin1988} (solid-lines). (b) shows the molar enthalpy ($\bar{H}_m$) for various weights (symbols) as a function of temperature.
  }
  \label{fig:wPtAu}
\end{figure}

An independent consistency check on the previous results can be done by comparing the chemical potential gradients ($\Delta \mu^\nu$) of each phase $\nu$.
Fig.\ref{fig:dMuPtAu}(b) plots the chemical potential gradients in each phase $\Delta \mu$ at various temperatures, where the values of the chemical potentials gradients are measured using the Widom test (see App.\ref{app:a}). For reference, a value obtained from the Gibbs-Duhem integration approach is also shown (dashed line) which can be obtained numerically via\cite{Mori}
\begin{align}
\frac{d \Delta \mu^{eq}}{dT} = \frac{\Delta h}{ T \Delta X} ,
\end{align}
where $\Delta X =  X^\alpha-X^\beta$ is the difference in concentration and $\Delta h = h^\alpha - h^\beta$, is the difference in enthalpy (per atom) between the two phases. Note, that the Gibbs-Duhem approach only allows us to get this curve up to a constant. Nevertheless, the shape of the curve can 

\begin{figure}[t]
  \centering
  \includegraphics[scale=0.220]{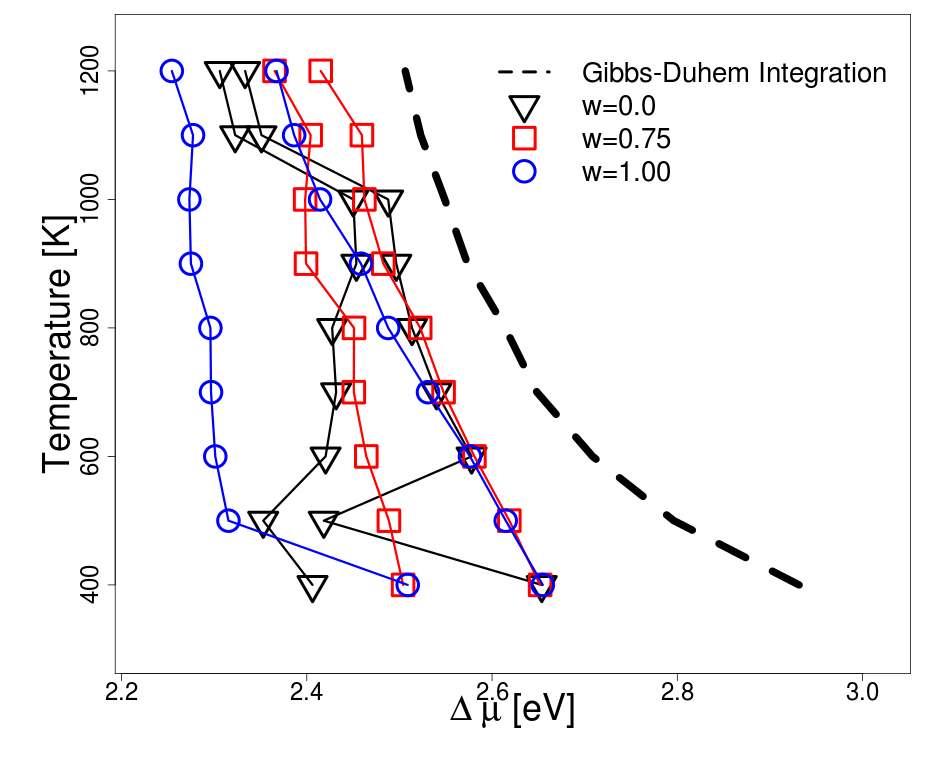}
  \caption{ 
   $\Delta \mu$-T plot (symbols) for different weights $w$ used in the predictor-corrector approach, compared against the Gibbs-Duhem integration method (dashed-lines);
For clarity only three sets of data are shown.
  }
  \label{fig:dMuPtAu}
\end{figure}

be contrasted with the values obtained using \abb. The case with $w=0.75$ reproduces a smooth curve similar to the Gibbs-Duhem approach. On the other hand, extreme values of the weight, i.e. $w=0$ or $w=1$, tend to show discontinuities in the $\Delta \mu$-T curves which are not expected while traversing the coexistence line of miscible alloy mixture. Note that \abb does not require additional methods in order to find a constant of integration, rather these values are found self-consistently. 

 As is customary in MC searches, acceptance rates based on MC moves need to be adjusted to be large enough to find a convergence state, but not too large to accept any spurious configuration. In flip attempts, the designated number of particle to be ``flipped'' is determined by choosing a random number of particles $n_{max}$ to be flipped within a given cell. This number is adjusted during the simulation roughly every 500 cycles, by increasing (or decreasing) the maximum number of particles to be flipped in order to arrive at a predetermined acceptance rate. It was found that acceptance rates that are larger than 30\% became problematic at larger temperatures. Hence in this work, we stick to a 20\% acceptance rate on the flip MC moves, which for 108 atoms cell correspond to a maximum of about 3 flips at the same time ($n_{max} \sim 3 $). Likewise, translational displacements ($\vec{r}_{max}$) and volume changes $\Delta V$ cutoffs used to maintain the isobaric-isothermal state are adjusted to arrive at a target acceptance rate below 50\% and 10\%, respectively. It is recommended that volume changes are not accepted too often as this can lead to instabilities in the search process\cite{frenkel}. 

Lastly, we point out that one can speed up the search process by evolving the translational degrees-of-freedom (DOFs), as well as volume/shape changes using molecular dynamics and a thermostat/barostat, instead of sampling these DOFs using a metropolis approach. Most results in this paper were reproduced using a hybrid MD/MC approach, except for the case of the smallest cells. In this case, the temperature fluctuations are limited by the size of the box, and the magnitude of such fluctuations grows as the inverse of the cell size\cite{lebowitz,mishin2016}. Thus, we found that as the temperature increases, the cells tend to melt at lower temperatures compared to the larger cell counterparts. On the other hand, a pure MC approach introduces temperature only through the metropolis acceptance criterion, and in this manner melting is 
avoided. Based on our simulations for the Au-Pt system, we recommend a pure MC approach  when sampling  small cells (N<200) and temperatures beyond 1000 K.

\subsection{Size Effects}
Fig.\ref{fig:PtAu} shows the effect of different cell sizes on the prediction of phase boundaries using \abb. The phase diagram for various system sizes using \abb  are shown in Fig.\ref{fig:PtAu}(a) and are compared against the values obtained by thermodynamic integration obtained from Ref.\cite{obrien2018} and experimental data \cite{vesnin1988}. All \abb results are obtained considering the predictor-corrector algorithm with the optimal weight of $w=0.75$.   It can be seen that for temperatures below T<900 K, the coexistence boundaries for various cell sizes are essentially identical. However, at larger temperatures the algorithm seems to find another solution in larger cells (N=500 atoms) distinct from the smaller systems. Inspection of the larger cells show that the Pt-rich phase starts to form an interface inside the cell (Fig. \ref{fig:SnapPtAu}). 

\begin{figure}[t]
  \centering
  \includegraphics[scale=0.220]{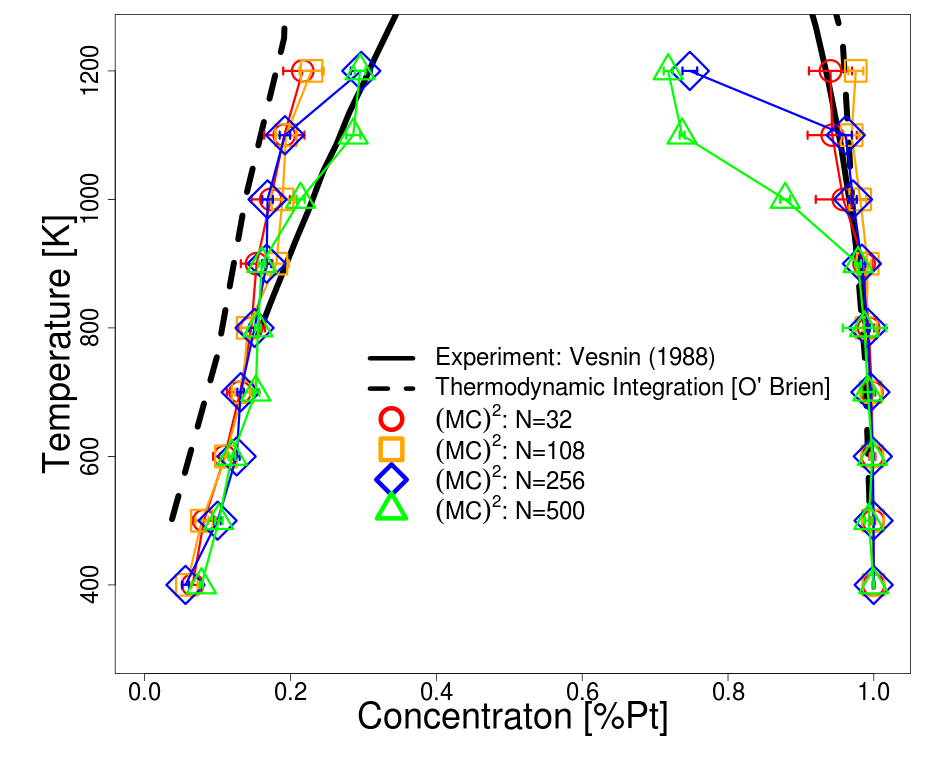} \\  \hspace{1cm}{\Large \textbf{(a)} } \\
  \includegraphics[scale=0.220]{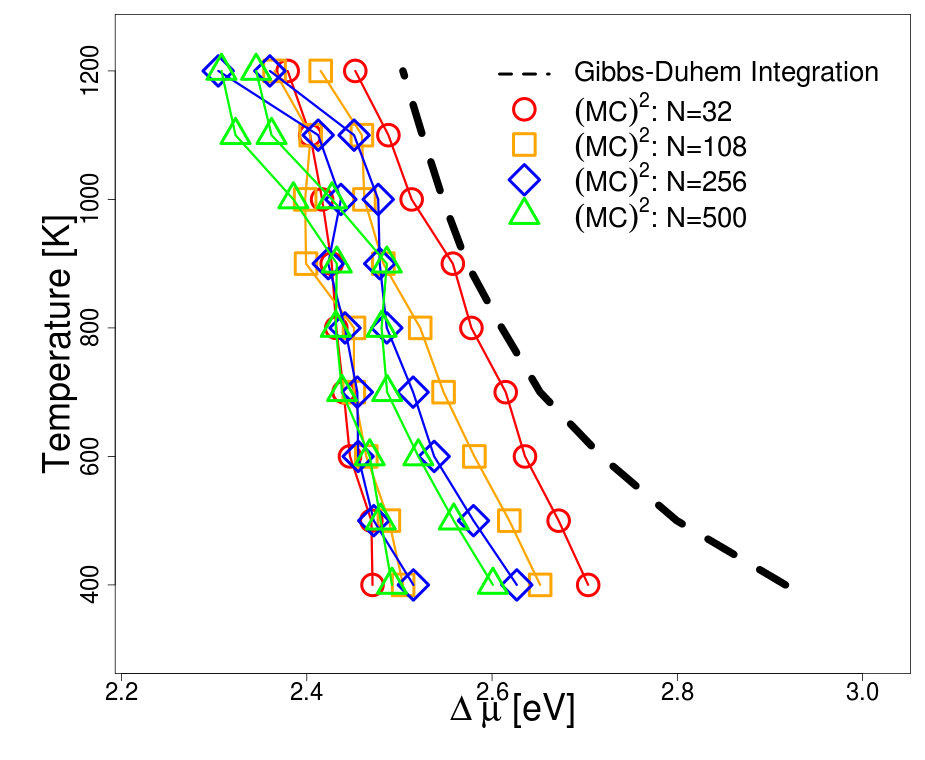} \\ \hspace{1cm}{\Large \textbf{(b)} } 
  \caption{ Effect of simulation size on \abb predictions.
    (a) shows the predicted phase diagram of the Au-Pt system (symbols), using various simulations sizes and $w=0.75$ in the predictor-corrector algorithm. Results are compared against Thermodynamic Integration calculation of O'Brian \textit{et al.}~\cite{obrien2018} (dashed-lines) and experimental data from Vesnin \textit{et al.}~\cite{vesnin1988} (solid-lines). 
    (b) shows the $\Delta \mu$-T plot (symbols) from simulations compared against the Gibbs-Duhem integration method (dashed-lines)
  }
  \label{fig:PtAu}
\end{figure}

\begin{figure}[t]
  \centering
  \begin{tabular}{cc}
    \multicolumn{2}{c}{108 atoms }\\
    \includegraphics[scale=0.41]{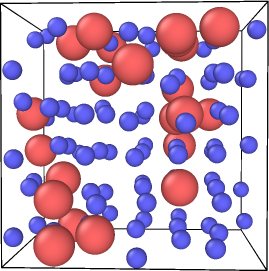} &   \includegraphics[scale=0.43]{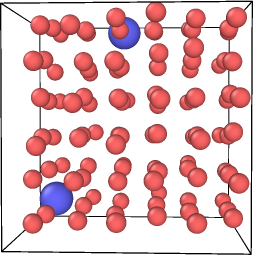}  \\
    Au-rich Phase  & Pt-rich Phase  \\
    \multicolumn{2}{c}{500 atoms }\\
    \includegraphics[scale=0.40]{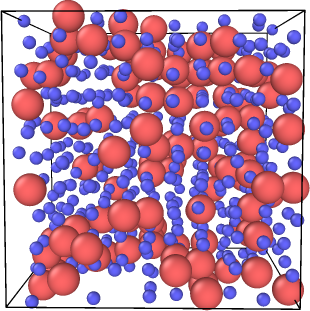}  &   \includegraphics[scale=0.41]{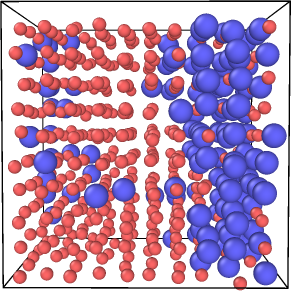} \\
    Au-rich Phase  & Pt-rich Phase  \\
  \end{tabular}
  \caption{
    Representative configurations of the Au-rich and Pt-rich phases at T=1100 K found for two cell sizes. For clarity, the atomic size of the most abundant species has been shrunk. Larger cells are prone to spinodal decomposition.
  }
  \label{fig:SnapPtAu}
\end{figure}

\begin{figure}[t]
  \centering
  \begin{tabular}{cc}
    \includegraphics[scale=0.220]{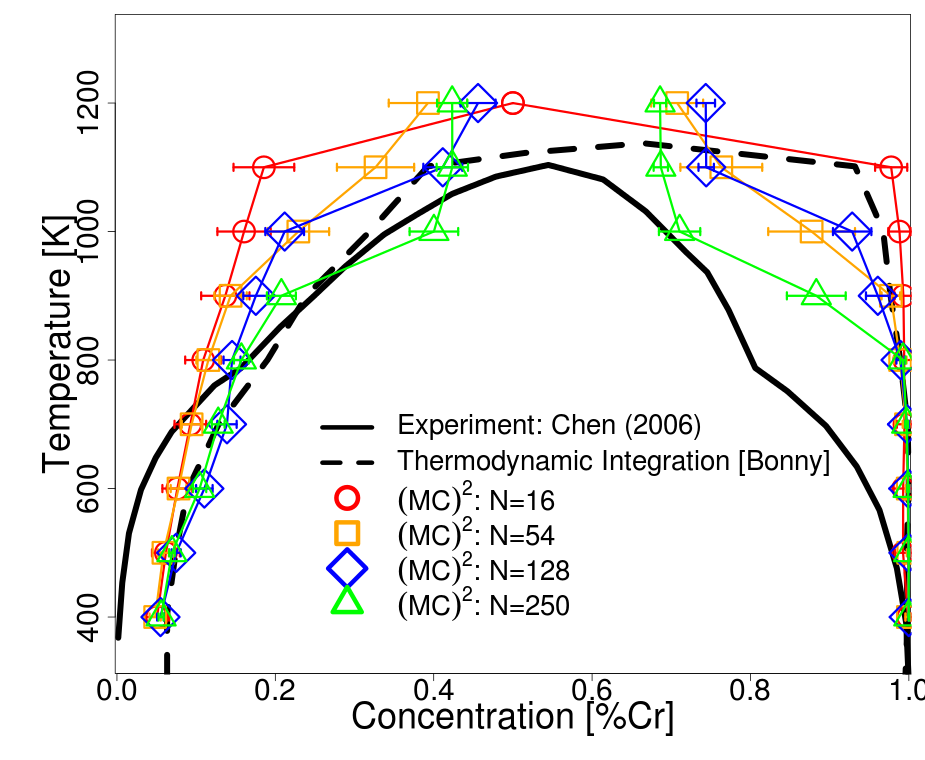} \\  \hspace{1cm}{\Large \textbf{(a)} } \\
    \includegraphics[scale=0.220]{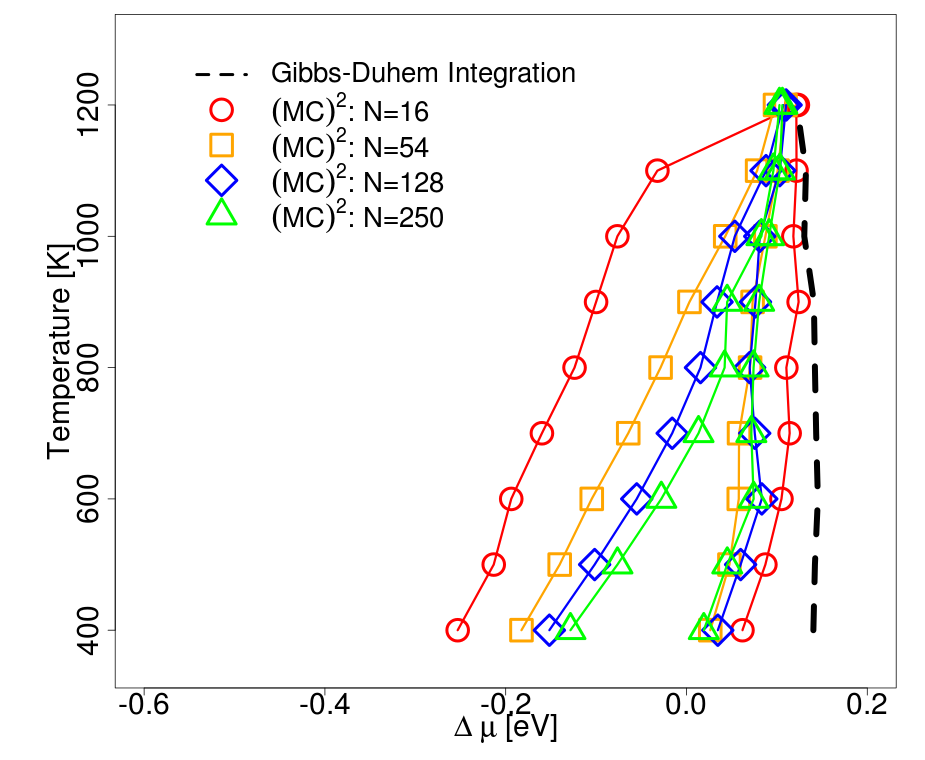} \\  \hspace{1cm}{\Large \textbf{(b)} } 
  \end{tabular}
  \caption[]{ Phase diagram of Fe-Cr system, predicted by \abb.
    (a) Calculated phase diagram of the Fe-Cr system (symbols) compared  against Thermodynamic Integration calculations of Bonny \textit{et al.}~\cite{bonny2011} (dashed-lines) and experimental data from Chen \textit{et al.}~\cite{Chen2006} (solid-lines). (b) $\Delta \mu$-T plot (symbols) compared against the Gibbs-Duhem integration method (dashed-lines)
  }
  \label{fig:FeCr}
\end{figure}

\begin{figure}[t]
  \centering
  \begin{tabular}{cc}
    \multicolumn{2}{c}{54 atoms }\\
    \includegraphics[scale=0.40]{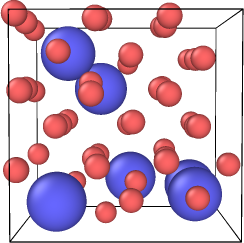}  &   \includegraphics[scale=0.41]{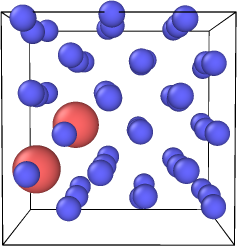} \\
    Fe-rich Phase  & Cr-rich Phase  \\
    \multicolumn{2}{c}{250 atoms }\\
    \includegraphics[scale=0.40]{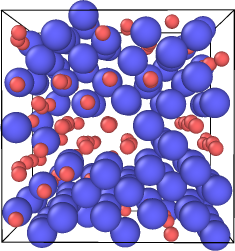}  &   \includegraphics[scale=0.41]{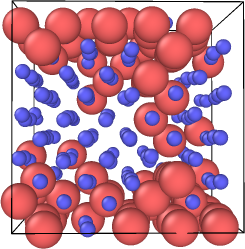} \\
    Fe-rich Phase  & Cr-rich Phase  \\
  \end{tabular}
  \caption{
    Representative configurations of the Fe-rich and Cr-rich phases at T=1100 K found for two cell sizes. For clarity, the atomic size of the most abundant species has been shrunk.
  }
  \label{fig:SnapFeCr}
\end{figure}

As discussed in the previous section, we can verify the \abb predictions of the equilibrium phase boundaries by measuring the difference in chemical potentials in each phase using the Widom test and comparing against an estimate via Gibbs-Duhem integration approach. These values are shown as a $\Delta \mu$-T plot shown in Fig.\ref{fig:PtAu}(b) where the chemical potentials obtained in \abb are in symbol for various cell sizes and dashed line corresponds to the Gibbs-Duhem estimate; the curves obtained using the Gibbs-Duhem integration are essentially identical for all system sizes therefore only a single curve is shown here for clarity.   

Cases where the coexistence lines in the phase diagram deviates from the main trend, also show a deviation in $\Delta \mu$-T with respect to the Gibbs-Duhem solution. Note that as the simulation cell increases, the chemical potential gradients for the two phases, $\Delta \mu^{\alpha}$ and $\Delta \mu^{\beta}$, get closer to one another. Such finite-size effects are expected to vanish in the thermodynamic limit ($N\rightarrow \infty$). 
The non-vanishing gap between these values is also consistent with the hysteresis seen in Semi-Grand Canonical Ensemble (SGCE) simulations where the transition $\Delta \mu$ observed depends on which side ($\alpha$ or $\beta$) the SGCE simulations is chosen as the starting point\cite{williams2006}. 

Consider now a Fe-Cr binary system which shows a solid-solid miscibility gap with a Fe-rich BCC solid solution phase and a Cr-rich BCC solid solution phase. Here, the inter-atomic potential, proposed by Bonny \textit{et al}, is used to recreate the phase diagram.  This potential was fitted to thermodynamic parameters and point-defect properties obtained from DFT calculations and experiments\cite{bonny2011}. 

Fig.\ref{fig:FeCr}(a) compares the $\text{MC}^2$ predictions to the thermodynamic integration performed by by Bonny \textit{et al} (dashed lines) and experimental data (solid lines).  Similar to the previous case, the values obtained below 900 K are in reasonable agreement with the values obtained using standard thermodynamic integration approach. Above this temperature, differences in the phase boundaries start to emerge. As in the Pt-Au case, the largest cells start to form interfaces within the cell (see Fig.\ref{fig:SnapFeCr}), and as a result the predicted coexistence boundaries  at higher temperatures are not valid in these cases. The chemical potential difference measured in the simulation (symbols)  are compared against the estimate from the Gibbs-Duhem integration (dashed-line)  in Fig.\ref{fig:FeCr}(b). Similar to the previous alloy, the gap between chemical potential gradients between the two phases becomes smaller as the system size increases. 

Near the top of the miscibility curve, an identity issue arises for some cell sizes where the concentration of the cells tend to flip occasionally.  The solutions of the molar fractions are bound by the number of atoms in the cells, that is, these solutions are discrete. As the concentration of the two cells become closer to each other, energetically equivalent configuration can be chosen spontaneously if the molar fractions and concentrations of the cells suddenly change identities. As the number of particle in the cell increases, the number of discrete solution increases and therefore the jumps that give rise to a change in identity become less likely. A similar effect has been also reported near the vicinity of the critical fluid-gas miscibility curve by the original Gibbs Ensemble approach.\cite{panagiotopoulos1992}

To summarize, in this section we find that the results obtained via \abb can reproduce the main features of phase diagrams obtained using  well established methods which compare the actual free energies of the two phases. However, as the size of the cells increases and the temperatures sampled become large (T>1000K), it is not possible to retain the lowest free energy solution with the current scheme. This is likely because the current implementation of the predictor-corrector approach only checks for a vanishing energy variation to the first order in mass variation, but does not check that such variations are positive definite to second order. Therefore, if cell sizes are large enough, they can individually undergo spinodal decomposition as shown in Figures~\ref{fig:SnapPtAu} and ~\ref{fig:FeCr}. A more stringent condition that also satisfies this condition can be explored in the future. Also, we emphasize that all simulations shown here were run independently from one another. It is feasible to also make use of thermodynamic information from adjacent temperature regimes, i.e. Gibbs-Duhem relation, in order to remain close to the coexistence line as is done in the approach of Kofke and co-workers\cite{kofke1993,Mori} and avoid falling into metastable solutions.
Such approach will be explored in future. Having shown a good correspondence with phase diagrams of binary alloys and the current limitation thereof,  next we explore the stability of a model quaternary alloy.

\subsection{Phase coexistence on a multicomponent alloy}

Reliable interatomic potentials beyond binaries are essentially non-existent. Here, we demonstrate the application  of our method to a model NiCoFeTi-alloy using the Zhou-Johnson potential~\cite{Johnson}. Previously, the deformation properties of this quaternary alloy in FCC and BCC structures have been explored by Rao \textit{et al}, showing good agreement in yield strength for similar complex concentrated alloys\cite{raoFCC,raoBCC}. In that work, optimal composition and structure for this quaternary alloy at zero temperature were found by sampling the composition space that maintained the stability of the lattice with respect to Bain deformation, elastic stresses and large shear deformations on the (111) planes for FCC and (110) planes for BCC. The result of this search found that the structures FCC-$Ni_{0.36}Co_{0.30}Fe_{0.16}Ti_{0.16}$\cite{raoFCC} and BCC-$Ni_{0.16}Co_{0.16}Fe_{0.366}Ti_{0.30}$\cite{raoBCC} satisfied the above conditions.

\begin{figure}
  \centering
  \begin{tabular}{cc}
  \multicolumn{2}{c}{\includegraphics[scale=.3]{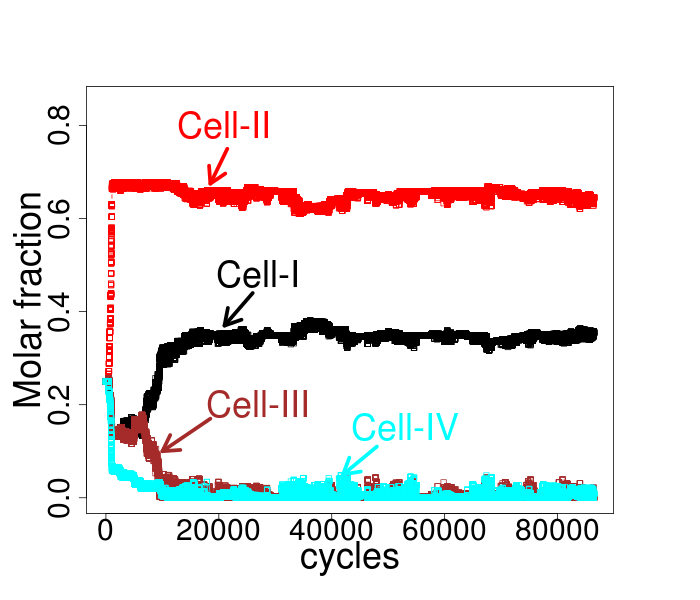}} \\
  \multicolumn{2}{c}{(a)} \\
  \includegraphics[scale=0.275]{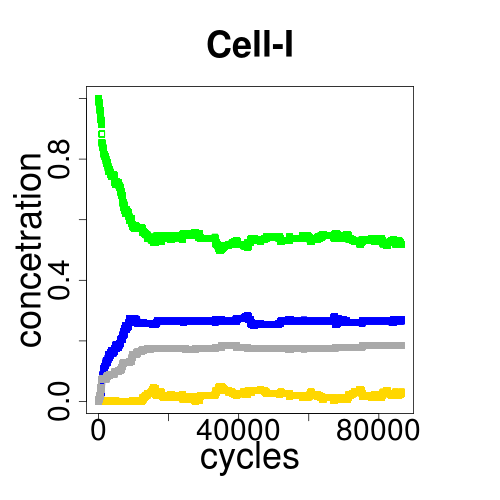} &  \includegraphics[scale=0.275]{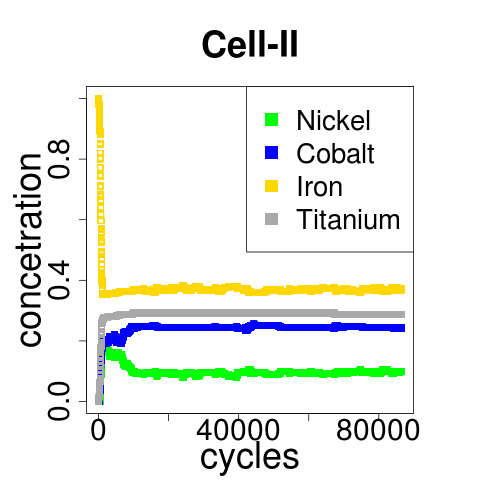} \\
  (b)  & (c)  \\
 \\
  \includegraphics[scale=0.35]{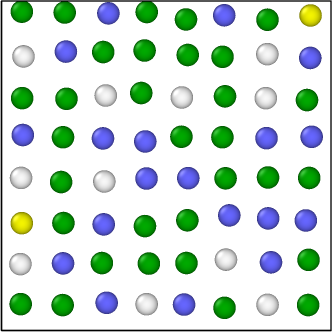} &  \includegraphics[scale=0.35]{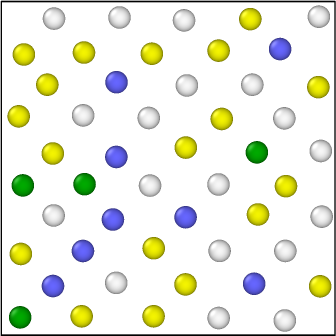} \\
  (d)  & (e) 
  \end{tabular}
  \caption[]{ \abb prediction for stable phases of the model NiCoFeTi system.
    (a) shows the progression of the Molar fractions vs the number of cycles using \abb on a four component systems (NiCoFeTi) starting from their pure elemental form at T=400K and zero pressure. (b)-(c) show the change in concentration of cells I and II and (d)-(e) project the corresponding atomic snapshots, where the different colors represent different elements.
  }
  \label{fig:QuatMol}
\end{figure}

In this work, we explore the same quaternary alloy, but instead we use the \abb method to find the most stable crystal structure and phase compositions. Using Eq.\ref{eq:phase_rule}, we assume that the maximum number of phases that can be achieved for this quaternary system are $\phi = C =4$, since both P and T will be used as independent intensive variables. Consider a coupled quaternary system starting with four elemental phases in their pure form: (I) FCC-Ni with 250 atoms, (II) BCC-Fe with 256 atoms, (III) HCP-Co with 250 atoms, and (IV) HCP-Ti with 250 atoms in each cell. To ensure that crystal structure of the different cells remain unchanged during as the algorithm progresses, only isotropic volume changes are introduced in the search process. 

Fig.\ref{fig:QuatMol}(a) shows the evolution of the molar fractions of each phase (cell) as the algorithm searches over the most stable configuration at T=400 K and zero pressure conditions. At this temperature, the algorithm finds two dominant phases corresponding to FCC and BCC structures, whereas the other phases considered are not energetically favored given that the molar fraction of cells III and IV (HCP structures) are negligible compared to cell-I (FCC) and cell-II (BCC). 

The  evolution in composition of each cell/phase during the search process is shown in Fig.\ref{fig:QuatMol}(b)-(c) for cells I and II respectively. 
After about 20000 cycles, the algorithm converges towards the structures: FCC-$Ni_{0.53}Co_{0.26}Fe_{0.02}Ti_{0.18}$ (Cell-I)and BCC-$Ni_{0.10}Co_{0.24}Fe_{0.36}Ti_{0.29}$ (Cell-II). A slice in two atomic layers along the (001) plane is shown in Fig.\ref{fig:QuatMol}(d)-(e) for the final compositions. Inspection of the FCC phase shows that Ti atoms prefer to avoid each other. This is expected in closest packed structures since Ti atoms have a larger atomic size ($\sim 20 \%$) relative to all the other elements\cite{miracle2001,Johnson}. On the other hand, the BCC structure appears to show a propensity for Fe-Ti bonds. These bonds are 
known to favor an ordered B2 phase at $Fe_{0.50}Ti_{0.50}$, yet for this composition, complete ordering of the phase B2 phase is not favored as 
there are significant traces of Co seen  in the final BCC structure. Notably, the BCC phases is close in composition to that found by Rao {\it et al}~\cite{raoFCC}, using mechanical stability of the lattice as the only criterion, while the FCC phase found by Rao {\it et al} is much less rich in Fe.\cite{raoBCC}. Nevertheless, the final concentrations depends on temperature as well.

Fig.\ref{fig:HEA} shows the effect of temperature on the final molar fractions and composition for all the dominant phases. Only the molar fraction for Cells I, II, and III are shown explicitly since the molar fraction for Cell-IV is significantly smaller than all the other cells considered over the whole temperature range. For this reason, Cells I, II and III are hereafter referred to as FCC phase, BCC phase, and HCP phase,  respectively. \abb shows that for this quaternary system, a BCC phase is expected to be energetically favored for the entire temperature range since its molar fraction shows the largest value. Moreover, a second FCC phase is also favored up to about $T \sim 900 K$, at which point a third HCP phase starts to emerge at the expense of the FCC phase

In order to visualize the phase compositions for this quaternary alloy, we project ``pseudo  phase diagrams'' for each of the elements making up the system (NiCoFeTi). This is shown in Fig.\ref{fig:HEA}(b), where the different symbols  indicate whether the system belongs the BCC-phase ({\LARGE $\circ$}), FCC-phase ({\LARGE $\diamond$}), or HCP-phase ({$\triangle$}). Note that below, (T<900K) the HCP is not shown since its molar fraction is negligible compared to the other phases. As expected, the concentration of each element in different phases are not independent, that is $X^{FCC}_{Ni}+X^{FCC}_{Co}+X^{FCC}_{Fe}+X^{FCC}_{Ti}= 1$, and similarly for all the other phases.

 \begin{figure}[h]
 
  \hspace{-1.25cm}\includegraphics[scale=0.31]{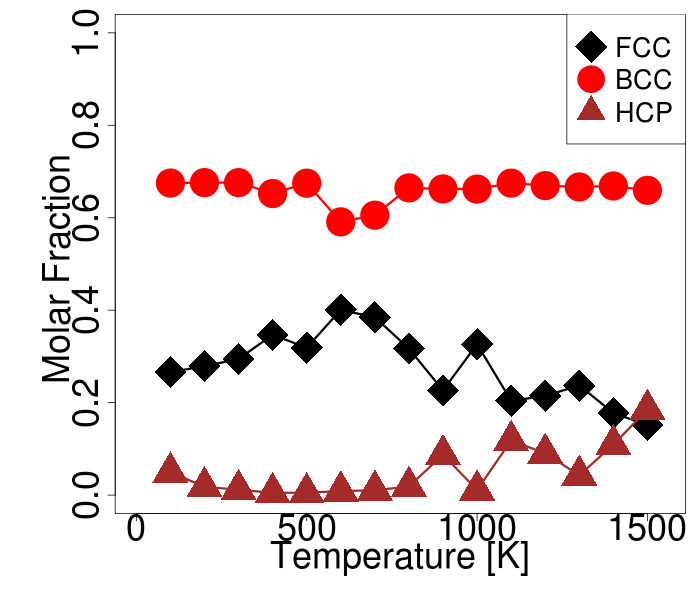} \\ {\Large \textbf{(a)} } \\
  \includegraphics[scale=0.26]{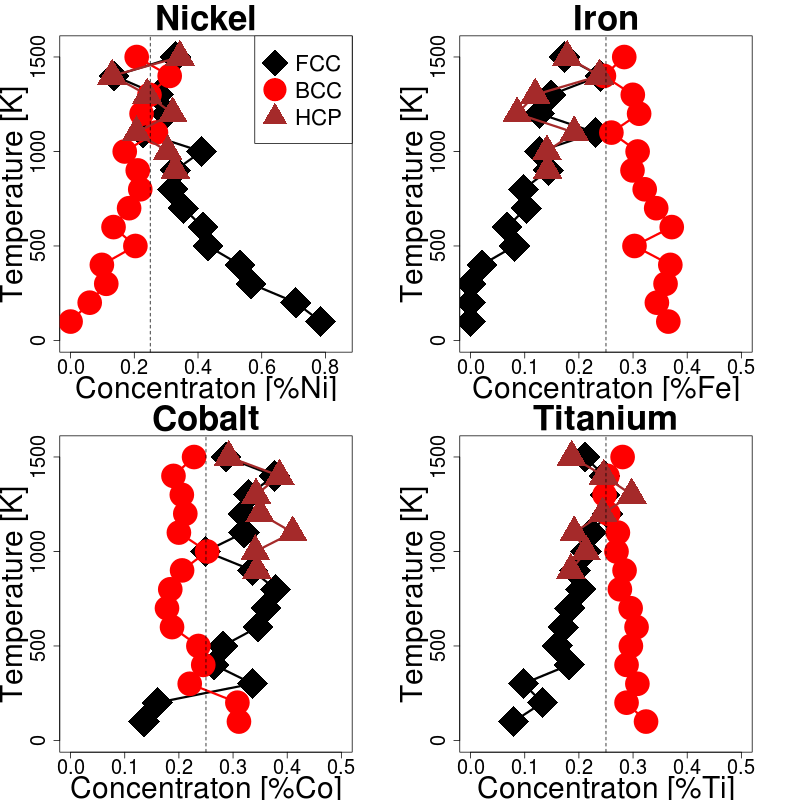} \\ {\Large \textbf{(b)} } \\
  \caption{ Stable phases of model NiCoFeTi alloy at different temperatures as predicted by \abb.
    (a) shows the equilibrium molar fractions vs temperature for the dominant phases considered in the NiCoFeTi alloy during the MC search. (b) 
    ``Pseudo phase diagrams'' showing the composition of every element vs temperatures for the three phases considered: FCC({\LARGE $\circ$}),  BCC ({\LARGE $\diamond$}) and HCP ({$\triangle$}); the dashed line is a reference at 0.25 
  }
  \label{fig:HEA}
\end{figure}

Inspection of the various pseudo-phase diagrams shows various pseudo-miscible gaps between the FCC , BCC,  and HCP lattice structures over all the elements in this system. Nickel shows the widest miscible gap starting from 80 \% Ni-rich phase in FCC, and 0\% Ni-rich BCC phase  at the lowest temperature sampled (T=100K). As the temperature increases the gap starts to rapidly close around T = 1000 K, at which point the Ni concentration on all the phases considered is about the same, i.e. 25 \%.  Iron and Titanium show similar trends and miscible critical temperatures ($T_c \sim 1000 K$), although the overall ranges in the concentration of the miscible gaps tend to be narrower compared to that of Nickel. Interestingly, Cobalt shows oscillation in concentration around the equiatomic composition throughout the whole temperature range.  As the temperature increases, a progression towards equiatomic compositions for all the phases, indicates that
the entropic terms start to contribute significantly to the free energies of mixing. Notice that above T>900 K, 
the concentration in the FCC and HCP phases trace very similar paths for all the elements, indicating that above this temperature these two crystal structures have similar energetics. This is also evidenced by the fact that the molar fraction of both of these phases appear to be anti-correlated in this temperature range.

The results obtained for the ``pseudo-phase diagrams'' in this multicomponent alloy will depend largely on the interatomic potentials \cite{Johnson}, and thus are valid to the extent of the reliability of the potential. The intent here is to show that the \abb formalism can be readily applied to multicomponent alloys to search for solid-solid phase coexistence without introducing additional complexity into the search process. Such approach can be used as a predictive tool for phase equilibria of multicomponent systems using \textit{first-principles} calculations\cite{rao2019}.

\section{summary}
\label{sec:summary}

We presented important improvements to the \abb method for modeling phase coexistence. The \abb method, is similar to Gibbs ensemble MC in the sense of using multiple simulation cells. While Gibbs ensemble technique has been a successful approach to study liquid-gas and liquid-liquid coexistence, its direct adaption to crystalline solids has been prohibitive because particle transfer between crystalline cells creates highly energetic point defects. In reference~\cite{rao2019}, Niu \textit{et al.} proposed a virtual mass transfer method, by randomly changing the chemical identity of species in each phase, while maintaining the overall composition via solving the mass balance equation, i.e. the lever rule. In this work, we first derived the most general acceptance criteria for \abb moves, starting from the NPT Gibbs ensemble for mixtures, and show that \abb samples a pseudo-Gibbs ensemble. The term ``pseudo" is chosen in accordance with previous literature and is meant to differentiate between the virtual mass transfer and actual insertion/deletion of particles. 

Next, we showed that relying on lever rule alone to search for the minimum free energy poses a practical shortcoming. Away from equilibrium, the solution to the mass balance equation is not unique. During the search for equilibrium, very different sets of concentrations and molar fractions can correspond to the same energy value. Consequently,  even though the energy may seem to  reach a plateau after several steps, the molar fractions and corresponding concentrations may not have converged to the proper equilibrium values. Therefore, it is necessary to define a stopping criterion for the simulation to ensure that in the end, the common tangent has been found. This is equivalent to the balance of chemical potentials at equilibrium. 

We then proposed an additional check on equilibrium via a predictor-corrector approach that aims to penalize solutions not satisfying the common-tangent criterion. We emphasize that our correction involves the \textit{difference} in chemical potential between species in different phases, rather than the chemical potentials themselves. This is necessary but not sufficient for equilibrium. The intent is to check whether or not the final phases are in equilibrium, and to steer the system towards  equilibrium. Constraints involving the individual chemical potentials in each phase guarantee equilibrium directly and will be pursued to further improve the method. 

The improved \abb was applied to binary and quaternary systems, using EAM potentials. Our findings highlighted the necessity to use the additional chemical potential constraint to avoid falling into nonequilibrium solutions. In case of binary alloys, the predictions were in good agreement with those of the standard thermodynamic integration. The thermodynamic integration approach is tedious and time consuming due to the fact that many simulations need to be performed around the neighborhood of the stable concentration of each phase for a given $\Delta \mu$\cite{williams2006}. \abb requires only a minimal number of coupled simulation at a given temperature, and as such lends itself easily towards the investigation of multicomponent system without excessive complication. We demonstrated this by applying \abb to a model quaternary system.

Lastly, we studied the role of simulation size on the predictions of the the phase boundaries and observed that larger cells are prone to spinodal decomposition. Another future direction is to incorporate conditions for the second derivatives of the free energy into the acceptance criteria to avoid decomposition within cells. However, we note that small cell sizes perform well, particularly at low to moderate temperatures. This is in fact advantageous when using the method with first principles calculations to predict stable phases of crystalline solids well below the solidus curve, which was the initial motivation for \abb.


\section{Acknowledgments}

This work is supported  by the Air Force Office of Scientific Research Grant FA9550-17-1-0168.  Computational resources were provided through the Ohio Supercomputer Center. We acknowledge helpful conversations with Wolfgang Windl and thank Michael Widom for valuable feedback.

\appendix

\section{Appendix}
\label{app:a}
\subsection{Widom test}

A method for calculating the difference between chemical potentials of solutes and solvents in the non-dilute solution
in the isothermal isobaric ensemble has been proposed by Frenkel \textit{et al}~\cite{frenkel1987}, which is based on an 
extension of Widom's potential distribution method~\cite{widom1982}. Assuming the contribution due to the ideal gas is negligible, the difference in the (excess) chemical potential can be obtained by using the following recipe: \\
~~\fbox {\textbf{Widom test particle method}}

  \begin{enumerate}
  \item  Attempt a virtual move, or flip: \\ 
    change a particle of species 1 into species 2
  \item  Compute the change in energy of the cell: $\Delta U$ 
  \item  Try many flips, but do not accept the move (otherwise composition changes) 
  \item Collect statistics \\
  \end{enumerate}
  The chemical potential difference is given by
  $$  \Delta \mu \equiv  \mu_{1} -\mu_{2}  = -k_B T \ln \langle \frac{N_{1}}{N_{2}+1} e^{-\frac{\Delta U}{k_B T}} \rangle $$
  where $N_1$ and $N_2$ are the number of species 1 and 2 respectively, and the $\langle \cdot \rangle$ is the ensemble average obtained from many flip attempts at various random sites occupied by species 1. Alternatively, changing species 2 to species 1  results in the negative of the above estimate. Large statistics on the above two average, ie  ($1\rightarrow2$) and ($2\rightarrow1$), can be used to estimate weighted average based on how frequently each species is flipped. This approach is used in the predictor-corrector method (see Section.\ref{sec:MCMC}.B) to measure the difference in chemical potential for each phase.

\subsection{Predictor-Corrector Formula}

\label{app:b}

In thermodynamic equilibrium, the local energy variation in mass needs to vanish to first order. This condition is demonstrated by
assuming that the free energy is a homogeneous function in the concentration of its chemical components\cite{lupis}, and expanding the energy changes with respect to the number of components to first order as : 

\begin{align}
  G^\alpha & (f^\alpha \cdot (n_1^\alpha + \delta n_1^\alpha),  f^\alpha  \cdot (n_2^\alpha + \delta n_2^\alpha),\cdots,
f^\alpha  \cdot (n_m^\alpha + \delta n_m^\alpha))  \nonumber  \\
   & =    f^\alpha G^\alpha ( n_1^\alpha + \delta n_1^\alpha ,n_2^\alpha + \delta n_2^\alpha, \cdots, n_m^\alpha + \delta n_m^\alpha) \nonumber \\
  & \approx  f^\alpha [ G^\alpha (n_1^\alpha ,n_2^\alpha,\cdots,n_m^\alpha) + \frac{\partial G^\alpha}{\partial n_1^\alpha} \delta n_1^\alpha + \frac{\partial G^\alpha}{\partial n_2^\alpha} \delta n_2^\alpha + \cdots ]  \nonumber \\
  & \approx  f^\alpha [ G^\alpha ( n_1^\alpha ,n_2^\alpha,\cdots,n_m^\alpha ) + ~\mu_1^\alpha~\delta n_1^\alpha~+~\mu_2^\alpha ~~\delta n_2^\alpha +
\cdots ] 
\end{align}
where the partial derivatives hold pressure and temperature constant. The variation in energy in phase-$\alpha$ is:
\begin{align}
\label{eq:dG}
  \delta G^\alpha =  f^{\alpha} [\mu_1^\alpha \delta n_1^\alpha + \mu_2^\alpha \delta n_2^\alpha + \cdots + \mu_m^\alpha \delta n_m^\alpha] \nonumber \\
\end{align}
and the variation in energy (for all phases) is then given by 
\begin{align}
  \sum_\nu^\phi \delta G^\nu =  \sum_\nu^\phi \sum_i^m f^{\nu} \mu_i^\nu \delta n_i^\nu 
\end{align}

Since at equilibrium, $\mu_i \equiv \mu_i^\alpha = \mu_i^\beta = \cdots  = \mu_i^\phi $, and the overall system is closed such that mass transport between the phases satisfies the lever rule ( $\sum_\nu f^\nu  \delta n_i^\nu = 0 ~ \forall~i~$). It follows that the above expression vanishes, viz
\begin{align}
  \sum^\phi_\nu \delta G^\nu =  \sum^m_{i } \mu_i \sum^\phi_\nu f^{\nu} \delta n_i^\nu = 0 
\end{align}

In the implementation of the \abb algorithm described in reference\cite{rao2019} did not check for the equality of chemical potentials directly.  Since mass is not physically transferred between phases, i.e. it is only mimicked,  it is not guaranteed that common tangent criterion is satisfied. Thus, to avoid falling into these solutions, a consistency check on the energy variations as a result of the virtual mass changes is proposed. Every time a particle of ``type j'' is replaced by another ``type i'' in phase $\alpha$, the variation in mass of each species is $\delta n^\alpha_i=+1$ and $\delta n^\alpha_j=-1$.  (See Eq.\ref{eq:dG}). Hence the energy change due to the change in chemistry ( $j \rightarrow i$) is given by 

\begin{align}
  \delta G^\alpha(j \rightarrow i)  =  f^{\alpha} [     (\mu_i^\alpha -  \mu_j^\alpha)   ]
\end{align}
Clearly if more than one replacement of the same kind ($j \rightarrow i$) occurs, the energy change is a multiple of this value, ie
\begin{align}
  \delta G^\alpha(j \rightarrow i)  =  f^{\alpha} \Delta \mu_{ij}^\alpha \delta n^\alpha_{ji}
\end{align}
where, $\Delta \mu_{ij} = \mu_i - \mu_j$ and $\delta n_{ji}$ is the number of lattice sites  where a replacement ($j \rightarrow i$) has occurred. 
After a flip move takes place, mass is transferred (at least conceptually) to another phase. 
This phase is chosen at random, from all the other possible phases, say phase-$\beta$. 
The change in chemistry in phase-$\beta$ that occurs as result of flipping particles ($j \rightarrow i$) in phase-$\alpha$ can be written as 
\begin{align}
\label{eq:dnbeta} 
\delta n^\beta_{ji} = -(f^\alpha/f^\beta) \delta n_{ji}^\alpha~\forall~(i,j)
\end{align}
while chemistry in all other phases remains unchanged. Since we know how much mass is transferred to and from each phase, this 
allows us to predict energy changes based on current molar fractions values, and gradients of the chemical potentials ($\Delta \mu_{ij}$).

An explicit method (Euler method) can be used to predict the energy change due to mass transfer between phases $\alpha$ and $\beta$, as a result of the replacements ($j \rightarrow i$), 

\begin{align}
  \tilde{G}^{predicted}_m =   f^\alpha G_m^\alpha + f^\beta G_m^\beta + f^\alpha \Delta \mu_{ij}^\alpha  \delta n_{ji}^\alpha  + f^\beta \Delta \mu_{ij}^\beta  \delta n_{ji}^\beta 
\end{align}

An implicit method (the trapezoidal rule) can be used to correct the prediction based on the slopes at the next step
\begin{eqnarray}
  \tilde{G}^{corrected}_m &  =  f^\alpha G_m^\alpha + f^\beta G_m^\beta  + \frac{\delta n_{ji}^\alpha}{2} (f^\alpha \Delta \mu_{ij}^\alpha  + f'^{\alpha} \Delta \mu'^\alpha_{ij})  \nonumber \\ & + \frac{\delta n_{ji}^\beta}{2} (f^\beta \Delta \mu_{ij}^\beta  + f'^{\beta} \Delta \mu'^\beta_{ij})  \nonumber \\
\end{eqnarray}
where the terms with primes denote quantities evaluated at the next step. Note, that the \textit{slopes} of the current and proposed steps are combined in the above expression in order to correct the prediction by the Euler method. It is reasonable to 
also combine actual energy measurement of the next proposed step with the predicted values for that energy.
To do so, a \textit{weighted} value of the \textit{measured} and \textit{predicted} energies can combined for the next step.
\begin{align}
  G_m'(w) =  (1-w)\cdot G'_m + w\cdot\tilde{G}^{corrected}_m
\end{align}
where $0 \leq w \leq 1$. This approach is similar to Kalman-filter method\cite{kalman} where measured values and predicted values are combined based on the uncertainties of their respective values, here the weight ($w$) is used as a tuning variable. 

The proposed trials are accepted or rejected through the usual Metropolis-like criterion by looking at the difference between the corrected-predicted value and the energy value at the previous step, ie 

\begin{align}
\label{eq:wmetro}
 min\{1,e^{-\Delta \tilde{G}_m(w)/{kT}}\} 
\end{align}
where $\Delta \tilde{G}_m(w)  = G_m'(w)-G_m$. It is worthwhile to expand this expression (using Eq.\ref{eq:dnbeta}) to arrive at a modified acceptance criterion, namely Eq.\ref{eq:Gw}, used throughout this manuscript.

\bibliographystyle{abbrv}
\bibliography{thebib}{}

\end{document}